\documentclass[12pt,preprint]{aastex}

\newcommand{\lapprox}{{\footnotesize $\buildrel < \over \sim \,$}}

\shorttitle{Dust around T~Cep}
\shortauthors{Guha Niyogi, Speck \& Onaka}

\begin{document}


\title{A temporal study of oxygen-rich pulsating variable AGB star, T~Cep: Investigation on dust formation and dust properties}


\author{Suklima Guha Niyogi\altaffilmark{1},
Angela K. Speck\altaffilmark{1} \& Takashi Onaka\altaffilmark{2} }
\altaffiltext{1}{Department of Physics \& Astronomy, University of Missouri,
Columbia, MO 65211}
\altaffiltext{2}{Department of Astronomy, Graduate School of Science,
The University of Tokyo, Bunkyo-ku, Tokyo 113-0033, Japan}
\email{sgz94@mail.missouri.edu}


\begin{abstract}

Pulsation is believed to be the leading cause of dusty mass loss from
Asymptotic Giant Branch (AGB) stars. We present a temporal study of T~Cep, a
long-period Mira variable, using seven \textit{ISO} SWS spectra, covering a
16-month period over a single pulsation cycle. The observed spectral dust features
change over the pulsation cycle of this Mira. In general, the overall apparent
changes in spectral features can be attributed to changes in the dust temperature,
resulting from the intrinsic pulsation cycle of the central star. However, not all
feature changes are so easily explained. Based on direct comparison with laboratory spectra
of several potential minerals, the dust is best explained by crystalline
iron-rich silicates. These findings contradict the currently favored dust formation hypotheses.

\end{abstract}



\keywords{stars: AGB and post-AGB
--- stars: individual (T~Cep)
--- (stars:) circumstellar matter
--- (ISM:) dust, extinction
--- infrared: stars}


\section{Introduction}

Low and intermediate mass stars (0.8--8 M$_\odot$) are major contributors to
galactic chemical evolution. During the asymptotic giant branch (AGB) phase of
their evolution, these stars lose a significant fraction of their mass through
slow, massive winds at a rate of
$10^{-8} < \dot{M} < 10^{-4}$\,M$_\odot$\,yr$^{-1}$
\citep{vanLoon-et-al-2005}.
This mass loss occurs as a result of stellar pulsation \citep{Habing-1996} followed by
acceleration of dust grains by radiation pressure \citep[e.g.][]{Gehrz-Woolf-1971, Hofner-et-al-2007}.
Pulsation levitates atmospheric gas from the stellar surface.
Then, as the outflowing gas cools, dust can condense in the outflow and
capture the momentum of the star's luminosity. Given that pulsation drives
mass loss and affects the temperature and density
distribution in the outflowing material, the effect of pulsation should be
manifested in the nature of the dust, forming in the circumstellar region.

The mineralogy (i.e. composition and lattice structure) of the dust grains is
determined by both the prevalent chemistry, and the physical characteristics
(temperature and density distribution) of the outflowing gas. In turn, the
chemistry is determined largely by the C/O ratio, and to some extent by the
elemental abundances in the stellar atmosphere (metallicity).
Convection currents inside the star transport newly-formed elements
(especially carbon) from the helium- and hydrogen-burning shells to the
surface of the star.
Since carbon monoxide (CO) is a very stable molecule and
forms very easily in such stellar atmospheres, carbon and oxygen atoms combine
into CO, leaving the more abundant element to dominate the chemistry. Stars
start their lives with cosmic C/O $\approx 0.4$ \citep{Asplund-et-al-2009} and thus have oxygen-rich
chemistry. In some cases, the dredge-up of carbon is efficient enough to raise
C/O ratio above unity making the star carbon-rich \citep[e.g.][]{Wood-1988}.
Consequently oxygen-rich AGB stars can have 0.4 \lapprox C/O $<$ 1.

In this paper, we investigate a single long-period Mira variable T~Cep as a
case study to determine how the mineralogy and morphology of
circumstellar dust varies with the pulsation cycle of the central star.
We present analyses of time-variations in the infrared spectra, and show how
the spectral dust features change over the variability cycle of this Mira.
In \S~2 we discuss the history of silicate mineralogy; in \S~3 we describe the
observations of T~Cep; in \S~4 we describe our analyses of the dust features,
while the results are given in \S~5.
In \S~6, we investigate the mineralogy of the dust and how it varies with time as
well as the effect of grain shapes in dust spectral features. 
The overall discussion is in \S~7, and the summary and
conclusions are given in \S~8.

\section{Previous studies: dust around oxygen-rich AGB stars}

Dust is very important in many astrophysical environments.
In particular, due to the interplay between dust and stellar photons in
driving mass loss, it is necessary to understand the nature of the dust around
O-rich AGB stars.
Dust grains absorb starlight and reemit photons according to their temperature
and optical properties, typically in infrared (IR) region.
Composition, lattice structure, grain size, and morphology
all affect the resulting IR spectrum, as do the temperature and density
distribution in the dust shell. We can compare laboratory spectra of different
dust species to astronomical observations in order to identify the
dust species present in circumstellar envelopes.

\subsection{Finding silicates}
\label{findsil}

In the late 1960s, while investigating deviations of stellar energy
distributions from blackbodies, \citet{Gillett-et-al-1968} discovered a
peak near 10\,$\mu$m in the spectra of four late-type, evolved, variable stars.
Shortly thereafter a 10\,$\mu$m absorption feature was discovered in the
interstellar medium \citep[ISM;][]{knacke-et-al-1969,hackwell-et-al-1970}.
Initially this feature was attributed to silicate minerals
\citep{Woolf-Ney-1969}, based on spectra of mixtures of crystalline silicate
species predicted to form by theoretical models \citep{Gaustad1963,Gilman1969}.
However, the laboratory spectra of crystalline silicate minerals showed more
structure within the feature than observed in the astronomical spectra
\citep[see][]{Woolf-1973,hs73}.
Subsequent comparison with natural glasses
\citep[obsidian and basaltic glass; from e.g.][]{Pollack-et-al-1973}
and with artificially disordered silicates \citep{Day1979,kh79}
showed that ``amorphous'' silicate was a better candidate for the
10\,$\mu$m feature than any crystalline silicate.

Since the "amorphous" silicate 10\,$\mu$m feature was first observed, it has been found to be almost ubiquitous being
found in observations of evolved stars \citep[e.g.][]{Speck-et-al-2000,casassus-et-al-2001};
many lines of sight through the interstellar medium in our own galaxy
\citep[e.g.][]{chiar-et-al-2007};
and in nearby and distant galaxies
\citep[e.g.][]{hao-et-al-2005}.
Further spectroscopic observations of
O-rich AGB stars have found that these objects show a diverse range of spectral features,
which are interpreted in terms of dust condensation hypotheses
\citep[e.g.][]{Little-Marenin-et-al-1990,Sloan-Price-1995,Speck-et-al-2000}.
In order to understand how IR dust spectra are classified and the
features are interpreted we need to discuss the currently-favored dust
formation hypotheses.

\subsection{Dust condensation sequence}
\label{dustcond}

Three competing dust formation mechanisms have been considered for
circumstellar environments:
(1) thermodynamic equilibrium condensation \citep{lf99};
(2) formation of chaotic solids in a supersaturated gas followed by annealing
\citep{Stencel-et-al-1990};
(3) formation of seed nuclei in a supersaturated gas, followed by mantle
growth \citep{gs99}. The latter should follow thermodynamic
equilibrium as long as density is high enough for gas-grain reactions to
occur.
Several observational studies support this thermodynamic condensation
sequence \citep{Dijkstra-et-al-2005,Blommaert-et-al-2007}, however, the
interpretation of the observational data is, in part, dictated
by the condensation sequence.
In 2, chaotic grains form with the bulk
composition of the gas, and then anneal if the temperature is high enough
\citep{Stencel-et-al-1990}.

\begin{figure}[t]
\begin{center}
{\includegraphics[width=0.6\textwidth]{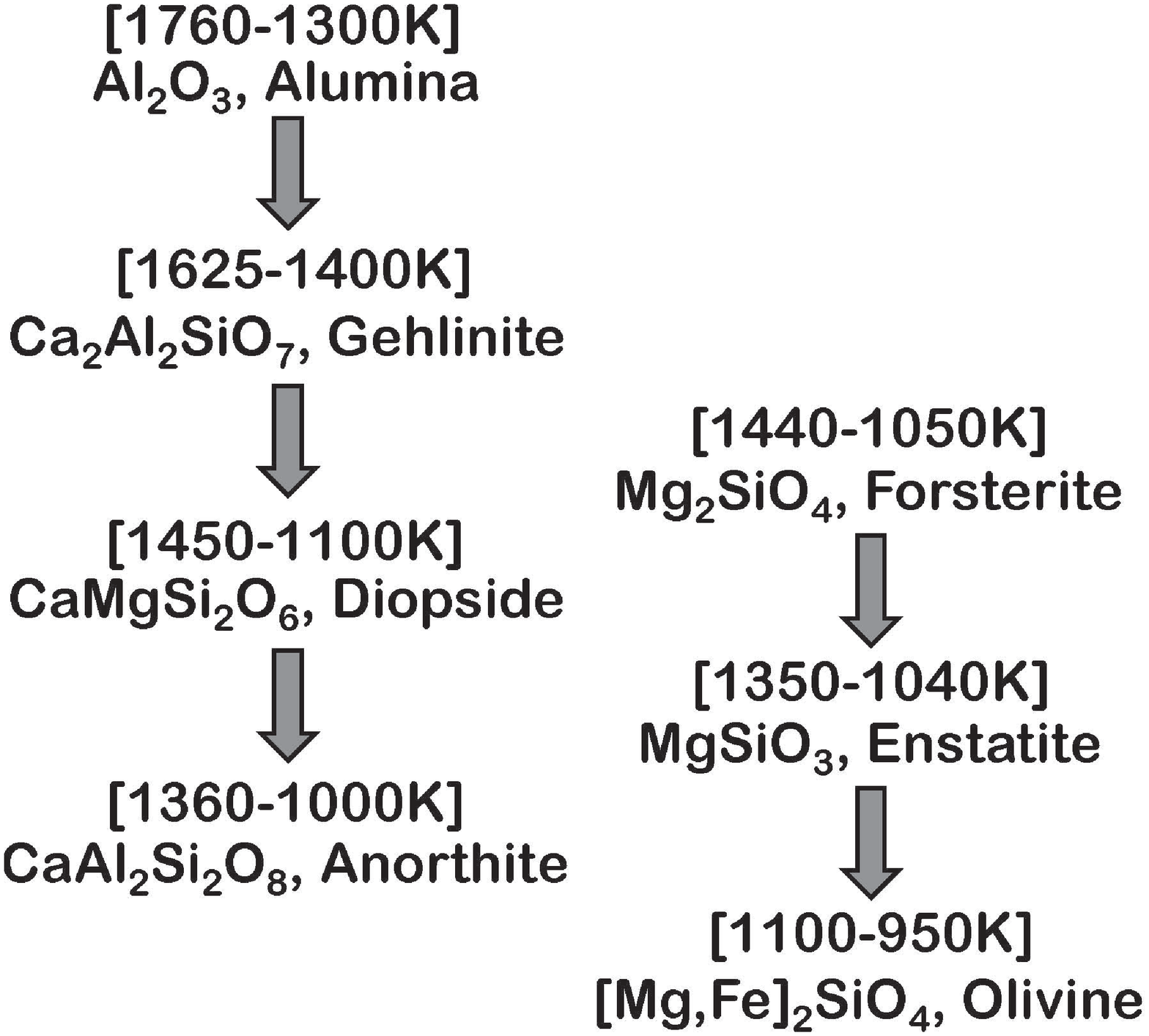}}
\caption{Predicted condensation sequence for O-rich environments from
\citet{Tielens1990} \citep[after][]{grossman72}.}
\label{fig:CondSeq2}
\end{center}
\end{figure}

The thermodynamic condensation sequence for a gas cooling down from a high
temperature is shown in Figure~\ref{fig:CondSeq2}
\citep[following][]{Tielens1990,grossman72}.
The condensation sequence starts with the formation of refractory
oxides (e.g. Al$_2$O$_3$) which, upon cooling, react with gaseous SiO, Ca and
Mg to form spinel (MgAl$_2$O$_4$, not shown in the figure), then to
melilite (ranging from gehlenite [Ca$_2$Al$_2$SiO$_7$] to \r{a}kermanite
[Ca$_2$MgSi$_2$O$_7$]) and then to diopside (CaMgSi$_2$O$_6$).
The conversion of diopside to anorthite (CaAl$_2$Si$_2$O$_8$) involves a
solid-solid reaction, and is therefore kinetically unlikely.

In this sequence, the more abundant Mg/Fe-rich silicates form as mantles on
the more refractory grains. Because of the relative abundance of
Mg and Fe when compared to Ca and Al, most of the silicon atoms condense as
the Mg-rich end member of the olivine family (forsterite: Mg$_2$SiO$_4$).
As the gas cools, the gaseous SiO converts
forsterite into the Mg-rich end member of the pyroxene
family (enstatite: MgSiO$_3$). Consequently, the MgSiO$_3$/Mg$_2$SiO$_4$ ratio increases
with the decreasing temperature.
At the pressure relevant for stellar outflows, gaseous Fe, may react
with enstatite to form
olivine with some iron ([Mg, Fe]$_2$SiO$_4$).
It is also predicted that Fe condense into metallic iron grains, but it
is not predicted that the Fe-rich end members of the olivine (fayalite: Fe$_2$SiO$_4$) or pyroxene series
(ferrosilite: FeSiO$_3$) will form.
Regardless of the precise mineral into which it condenses, iron is expected
to be in solid phase because it is strongly depleted from the gas phase \citep{Okada-et-al-2009}.

Stellar winds are commonly assumed to be dust-driven in circumstellar
environments of evolved stars \citep{Ferrarotti-Gail-2006}.
Opacity of dusty materials is an
important parameter for the wind-driving mechanism.
If iron is absent from dust, a stellar wind cannot be
driven because of the transparency of Mg-rich silicates.
Iron has often been included in dust grains in order to increase
the near-IR absorption efficiency  \citep{Tielens-et-al-1998}.
However, the precise effect of the iron
depends on how it is incorporated, i.e. as an integral part of the silicate
(e.g. in  amorphous [Mg,Fe]$_2$SiO$_4$ silicate) or as metallic inclusions.
For Mg-rich silicates with metallic iron inclusions, their absorption
efficiency is so high that radiative heating leads to grain evaporation and only a
limited amount of dust can survive \citep{woitke-06}.

For solar metallicity, Mg and Fe are approximately equally abundant in
O-rich circumstellar environment  \citep{lf99}.
In these dust condensation sequences, the condensation temperatures are
dependent on the gas pressure. However, for the pressures relevant to the
typical AGB outflow, the silicate formation temperatures are above the assumed
glass transition temperature of 1000\,K, implying that crystalline dust should
form.


\subsection{Spectral classification based on dust features}
\label{specclas}

Various attempts have been made to classify the observed dust features of
O-rich evolved stars \citep[e.g.][]{Little-Marenin-et-al-1990,Sloan-Price-1995,Speck-et-al-2000}.
All these authors have classified the observed spectra into groups according
to the shape of the dust feature, which reflects a progression from a broad
9--12\,$\mu$m
feature to the classic narrow 10\,$\mu$m silicate feature. This progression of
the spectral features is expected to represent the evolution of the dust from
the early forming refractory amorphous oxides (the broad feature) to the dominance of
magnesium-rich amorphous silicates (the narrow 10\,$\mu$m feature) as described in
\S~\ref{dustcond}.

The most recent version of IR spectral class based on silicate emission
is given by \citet{Sloan-et-al-2003}. They classify the silicate dust sequence
by taking the flux ratios at 10, 11 and 12\,$\mu$m.
They divided the sequence into eight segments, labeled SE1--SE8
(SE = silicate emission). Classes SE1--SE3 are expected to correspond to
low-contrast alumina-rich amorphous dust seen in evolved stars losing mass at
low rates and have optically thinner shells. Moving up the sequence,
classes SE3--SE6 show structured silicate
emission, with features at 10 and 11\,$\mu$m. The upper end of the silicate
dust sequences (SE6--SE8) consist of sources with the
classic silicate emission feature
believed to be produced by amorphous silicate grains. These sources have
optically thicker shells and higher mass-loss rates than sources at the other
end of the sequence.

In addition to the SE\# classification, based on the features in the
9--12\,$\mu$m range, another mid-IR feature at $\sim$13\,$\mu$m is sometimes
present. Approximately 50\% of O-rich AGB stars exhibit this feature and it has
caused of some controversy.
The so-called ``13\,$\mu$m feature'' was discovered in the \textit{IRAS} LRS spectra
\citep{Vardya-et-al-1986} and does not lend itself to the standard IR feature
classification system. It has been found in all SE classes
\citep{Speck-et-al-2000, Sloan-et-al-2003}.
Furthermore, the 13\,$\mu$m feature is associated with semiregular variables
(SRs) rather than Miras or red supergiants
\citep{Sloan-et-al-1996,Speck-et-al-2000}.
Many minerals have been proposed by various authors \citep[e.g.][]{Sloan-et-al-2003,DePew-et-al-2006}
as the likely carrier of this feature, including corundum (crystalline form of Al$_2$O$_3$),
spinel (MgAl$_2$O$_4$) and silica (SiO$_2$).
Both corundum and spinel are a predicted condensates,
presolar examples of which have been found in meteorites \citep[e.g.][and references therein]{cn04}.
However, spinel has been ruled out as the carrier of the 13\,$\mu$m feature  because it also shows a strong feature at 17\,$\mu$m (irrespective of relative abundance of spinel), and this feature does not occur in the spectra of
AGB stars that exhibit the 13\,$\mu$m
\citep[e.g.][]{Sloan-et-al-2003, Heras-Hony-2005, DePew-et-al-2006}.
Furthermore, \citet{DePew-et-al-2006} showed that only spherical grains of
spinel or corundum give rise to a narrow 13\,$\mu$m feature; other grain shape
distributions exhibit broader features at longer wavelengths.

One of the most exciting recent findings in cosmic dust studies was the
discovery of crystalline silicate dust by the Infrared Space Observatory
\citep[\textit{ISO};][]{Kessler-et-al-1996}
Short Wavelength Spectrometer \citep[SWS;][]{deGraauw-et-al-1996}.
Since then, crystalline silicates have been found in many
astrophysical environments including the solar system and extrasolar planetary
systems \citep[e.g.][and references therein]{Crovisier-et-al-1996,Mann-et-al-2006},
the circumstellar regions
of both young stellar objects \citep{Waelkens-et-al-1996}, and evolved stars
\citep{waters96, Molster-et-al-2002-a},
and in other galaxies \citep{Spoon-et-al-2006}.
These crystalline silicates were first observed around evolved stars with very
high mass-loss rates, leading to the inference that crystal formation
requires such conditions \citep[e.g.][]{cami98}.
However, recent studies on occurrence of crystalline silicates and their
distribution with evolution suggest that crystalline silicates preferentially
occur around less-evolved, low mass-loss rate evolved stars \citep[e.g.][]{Sloan-et-al-2010, McDonald-et-al-2011}.

The classical amorphous silicates have broad, smooth spectral features
at $\sim$10 and 18\,$\mu$m, with no diagnostic features at longer wavelengths;
that is not true for crystalline silicates. A crystalline
mineralogy would manifest itself more clearly at long IR wavelengths.
\citet{Molster-et-al-2002-a, Molster-et-al-2002-b, Molster-et-al-2002-c}
in their series of three papers studied dust
mineralogy of several oxygen-rich (post-)AGB stars using \textit{ISO} SWS data.
By comparing the observational spectra with laboratory data, these authors concluded presence of
Mg-rich crystalline silicates (both olivines and pyroxenes).
However, the carrier of two of the observed far-IR features at $\sim$20 and 32\,$\mu$m, need re-investigation.

Various authors \citep[e.g.][and references therein]{Begemann-et-al-1996}
pointed to some connection between iron-rich (and possibly Mg-rich) oxide and
an excess emission in the 19--20\,$\mu$m range. Shortly thereafter, the 19.5\,$\mu$m feature
was attributed to $\rm Mg_{0.1}Fe_{0.9}O$ independently by \citet{Posch-et-al-2002} and \citet{Cami-2003}.
Following this study, \citet{Lebzelter-et-al-2006} also attributed the same mineral for 20\,$\mu$m feature,
while studying the mid-IR dust features of AGB stars in the globular clusters 47 Tuc.
However, the occurrence of circumstellar $\rm Mg_{0.1}Fe_{0.9}O$ is not expected
from thermodynamic models \citep[e.g.][]{lf99} and the low stability temperature of
Mg-Fe oxides suggests that these minerals should not form directly from the gas phase.
Consequently a large abundance of $\rm Mg_{0.1}Fe_{0.9}O$ is unlikely.
Confirmation of this attribution is further hampered by the fact that
Mg-Fe oxides exhibit only one resonance vibrational band in the IR region, which precludes
verification of this composition by matching further features.
A recent \textit{Spitzer Space Telescope} \citep{Werner-et-al-2004} Infrared Spectrograph
\citep[IRS;][]{Houck-et-al-2004} study on globular cluster evolved stars
also suggested that FeO is a good candidate for this $\sim$20\,$\mu$m feature \citep{McDonald-et-al-2010}.
In addition their modeling suggests that metallic iron grains are present around
O-rich AGB stars, where it is observationally manifested as a featureless mid-IR excess.

Another interesting far-IR feature at $\sim$32\,$\mu$m was studied by
\citet{Molster-2000}, and tentatively attributed to crystalline diopside (CaMgSi$_2$O$_6$, a pyroxene).
Two recent independent studies of the spectrum of the AGB star
RX Lac, attribute the $\sim$32\,$\mu$m feature to two different minerals:
diopside \citep{Hony-et-al-2009} and fayalite \citep{Pitman-et-al-2010}.
Diopside exhibits a prominent ``25\,$\mu$m feature'' which is
absent in the spectrum of RX Lac, whereas, all fayalite spectral features are present.
\citet{Pitman-et-al-2010} studied the dust features of 4 O-rich AGB stars
from SE1 class (RX Lac, T Cep, T Cet, R Hya) using \textit{ISO} SWS data.
Using new laboratory data for crystalline olivines, they showed that the spectral
features for Mg$_2$SiO$_4$ are not present in the observed spectra.
Rather the authors found that
the peak positions and widths of the spectral features of these 4 stars match
better with non-endmember iron-rich silicate (Mg$_{0.18}$Fe$_{1.82}$SiO$_4$).
The authors concluded that this result is unexpected, but not entirely out
of the question, since it has been long assumed that silicates in space
include iron in order to account for their opacities.
Moreover, \citet{Tielens-et-al-1998} found iron-bearing dust to be more refractory than
typical silicates, and therefore more likely to survive, which may explain the \citet{Pitman-et-al-2010} result.
In addition to these spectroscopic studies, recent work on presolar silicate
grains from meteorites also suggests that there is more iron in silicate grains
around AGB stars, than our current models allow \citep[e.g.][]{stroud08, bose10}.

\section{Observations of T Cep}

The O-rich AGB star T~Cep, was discovered by Ceraski in 1878, has been classified as spectral
type
M5.5e - M8.8e \citep{Onaka-et-al-1999} and infrared spectral class of SE1
(i.e. low-contrast broad dust feature/low mass-loss rate) with a 13\,$\mu$m feature
\citep{Sloan-et-al-2003}. Its estimated distance is 210\,pc
from the HIPPARCOS catalogue \citep{Perryman-et-al-1997}.

T~Cep was observed using \textit{ISO} SWS for seven times in a 16-month period
(from August 1996 to December 1997)
covering one full stellar variability period.
The fully processed post-pipeline spectral data were acquired from an
online atlas associated with \citet{Sloan-et-al-2003}.
Detailed data reduction information is available from the atlas webpages\footnote{http://isc.astro.cornell.edu/~sloan/library/swsatlas/atlas.html}.
Dates of the observations and TDT numbers are listed in
Table~\ref{tab:lightcurve}, and all seven flux-calibrated spectra are shown in
Figure~\ref{fig:TCep-all7}
(seven observations are designated by sequential numbers).

\begin{figure}[t]
\begin{center}
{\includegraphics[width=0.75\textwidth]{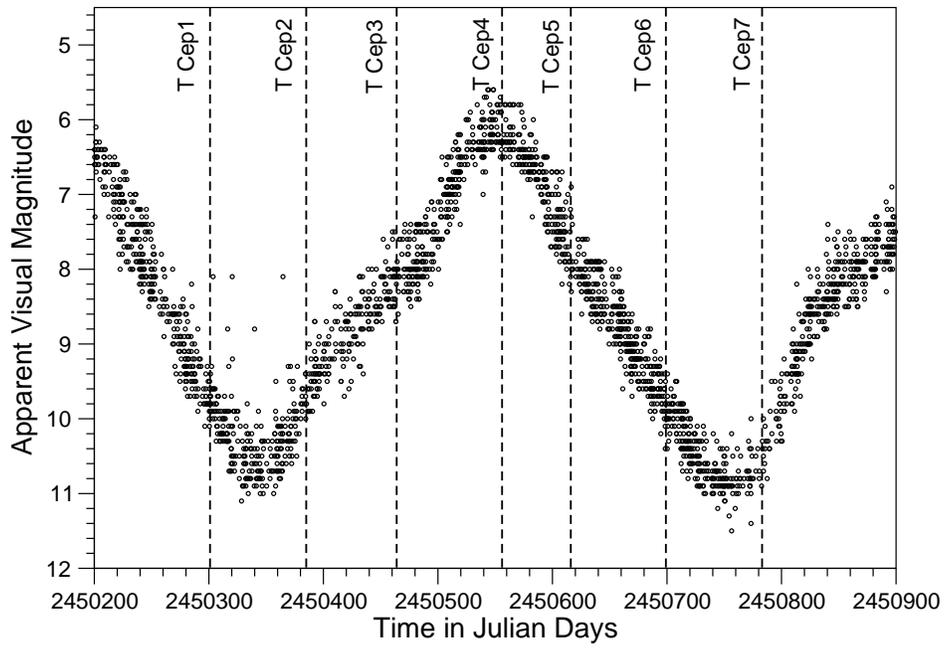}}
\caption{Light curve of T~Cep over a 16 month period.
The seven \textit{ISO} SWS observations are indicated by dashed straight lines.}
\label{fig:lightcurve-TCep-bw}
\end{center}
\end{figure}

In addition to the IR spectral data, we have also acquired light curve data
of T~Cep from American Association of Variable Star Observers (AAVSO)
database%
\footnote{http://www.aavso.org/}, which is shown in
Figure~\ref{fig:lightcurve-TCep-bw}. The positions in the
lightcurve at which \textit{ISO} SWS data was taken, are indicated by dashed straight lines. The apparent visual magnitudes at each \textit{ISO} SWS observation are listed in Table~\ref{tab:lightcurve}. According to these light curve data from AAVSO, the estimated pulsation period of T~Cep is 388 days, but the period shows evidence of variation. \citet{Castelaz-et-al-2000} reported the pulsation period is of 398 days, whereas, \citet{Weigelt-et-al-2003} showed the period is of 382 days. These variation of mean pulsation period of T~Cep is confirmed to be real by \citet{Isles-Saw-1989} and these authors also reported that T~Cep vary slowly in amplitude over the time. Also, an unusual secondary variation in T~Cep had been reported by \citet{Marsakova-Andronov-2000}, using the observation from AFOEV\footnote{ftp://cdsarc.u-strasbg.fr/pub/afoev} and VSOLJ\footnote{http://www.kusastro.kyoto-u.ac.jp} databases \citep{Schweitzer-1990,Nogami-1998}, obtained over 75-years. These authors also reported significant variation of amplitude and asymmetry over the pulsation cycle, and they indicated the possibility of interference of pulsations with two periods to explain this unusual variation. Similar findings for other several long period Mira variables are also  reported by several authors \citep[e.g.][]{Wood-Zarro-1981,Zijlstra-Bedding-2002,Benitez-Vargas-2002,Zijlstra-et-al-2004}.

\section{Analysing the dust spectral features of T~Cep}
\label{analysis}

In order to investigate the potential effects of the pulsation cycle on dust
formation, we need to consider what contributes to the observed spectra.
The spectrum,
$F_\lambda$, can be interpreted as a product of the underlying
continuum and an extinction efficiency factor ($Q_\lambda$) for the
entire spectrum,

\begin{equation}
\label{eqsimp}
F_\lambda = C \times Q_\lambda \times B_\lambda(T)
\end{equation}

\noindent
where $B_\lambda(T)$ is the Planck function for a black body of
temperature $T$, $Q_\lambda$ is a composite value including
contributions from all dust grains of various sizes, shapes,
crystallinities and compositions, and $C$ is a scale factor that depends
on the number of dust particles, their geometric cross section and the
distance to the star. In reality the spectrum should be represented
by:

\begin{equation}
\label{eqsum}
F_\lambda = \sum^{n,m}_{i=1, j=1}C_j \times Q_{\lambda,j} \times B_{\lambda,i}(T_i) = \sum^{n,m}_{i=1, j=1} C_j \times Q_j \times B_i
\end{equation}

\noindent
where each $B_i$ represents a single dust (or stellar) temperature
black body (of which there are $n$ in total), each $Q_j$ represents
the extinction efficiency for a single grain type as defined by its
size, shape, composition and crystal structure, and each $C_j$
represents the scale factor for a single grain type (of which there
are $m$ in total).

For T~Cep, which is a low mass-loss rate AGB star and exhibits low-contrast
spectral features, the spectrum is dominated by stellar photons.
Furthermore, since T~Cep has an optically thin dust shell
\citep[see e.g.][]{Onaka-et-al-1999, Sloan-et-al-2003}, its dust
spectrum should be dominated by the hottest and densest part of the dust
shell, which is essentially the inner dust radius.
Therefore we can simplify Eq.~\ref{eqsum} as:

\begin{equation}
\label{eqsum1}
F_\lambda = B_{stellar}(\lambda,T)+B_{dust}Q_{dust}
\end{equation}

\noindent
where $B_{dust}$ is the Planck curve for the inner dust temperature
($T_{dust}$), and $Q_{dust}$ is the emission efficiency for the innermost
dust grains. This assumption is an approximation to the dust continuum,
but the exact continuum does not significantly alter the subsequent analysis.

\begin{figure}[t]
\begin{center}
{\includegraphics[width=0.9\textwidth]{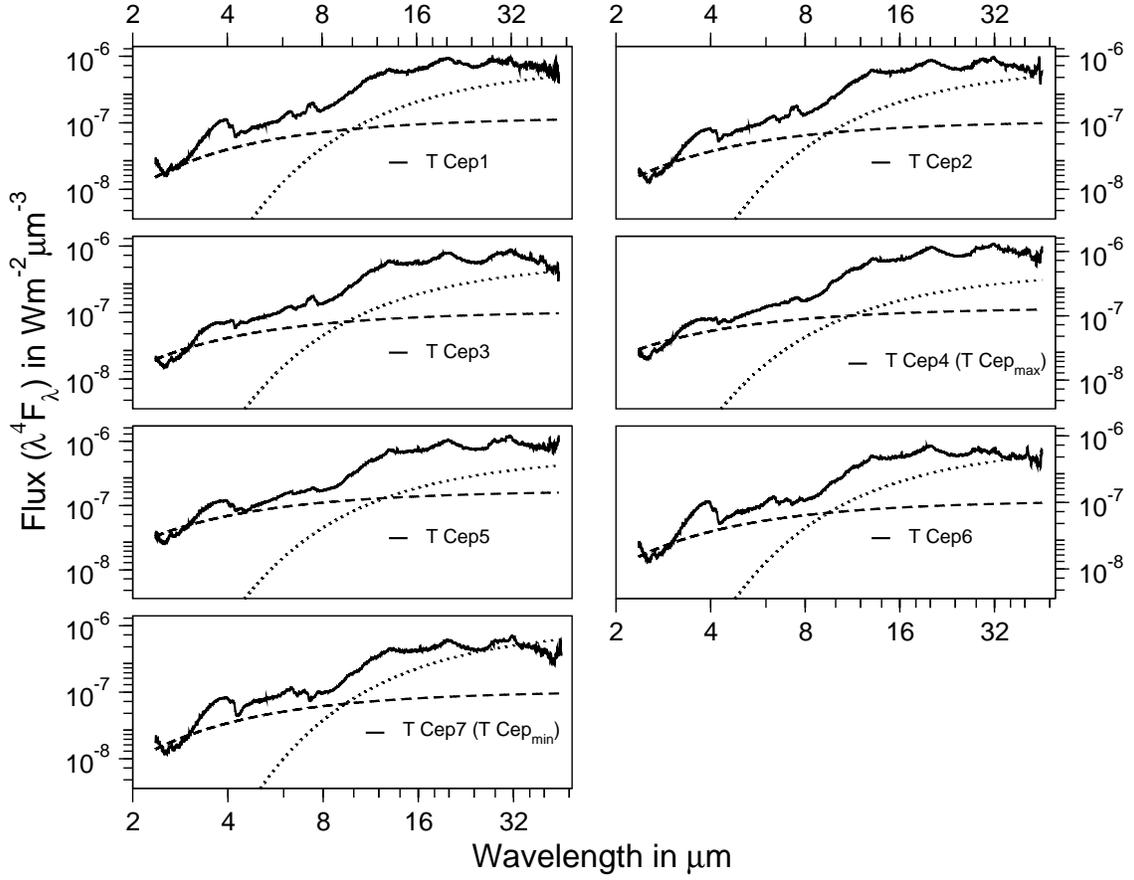}}
\caption{ \textit{ISO} SWS spectra of T Cep together with
both stellar  and dust continua.
{\em solid lines} are the observed spectra;
{\em dashed lines} are fitted stellar blackbody continua;
{\em dotted lines} are dust blackbody continua (see text for details).
The estimated temperatures of the stellar blackbodies ($T_{\rm eff}$) and the
fitted temperatures for the dust blackbodies ($T_{\rm dust}$) are listed in
Table~\ref{tab:lightcurve}.}
\label{fig:TCep-all7}
\end{center}
\end{figure}

Following Eq.~\ref{eqsum1} we can subtract the stellar contribution as
approximated by an appropriate blackbody, leaving only the contribution
from the dust shell.
The stellar spectral energy distribution (SED) is assumed to
be reasonably well simulated by a blackbody in the 2500--3500\,K range. The precise temperature used
in our analysis, was estimated from the spectral type and its
variability over time. For the spectral observation closest to maximum light
(T~Cep4; hereafter T~Cep$_{\rm max}$),
we use a temperature of 3347\,K,
determined from the spectral type of M5.5 \citep{Perrin-et-al-1998}; similarly
at minimum light (T~Cep7; hereafter T~Cep$_{\rm min}$) we used a temperature of
2566\,K, determined
from the spectral type of M8.8, by extrapolating the data from
\citet{Perrin-et-al-1998}. For the intermediate observations we linearly
interpolated between these minimum and maximum temperatures, relating to their
apparent changes in visual magnitude. Table~\ref{tab:lightcurve}  lists
the estimated stellar blackbody temperatures ($T_{\rm eff}$) for all the seven
observations. Figure~\ref{fig:TCep-all7} includes the original \textit{ISO} spectra (in solid lines) together
with the assumed stellar blackbodies (in dashed lines) in logarithmic scale
(normalized at 3.0\,$\mu$m, averaging over the range from 2.98--3.02\,$\mu$m)\footnote{%
while there may be molecular spectral features due to water within this range, visual inspection of the spectra shows these to be negligible, and the averaging over a range of wavelengths mitigates any potential problem. Furthermore since we are simply scaling a blackbody whose temperature is determined independent of the spectrum, small errors in scaling factor do not effect our results.}.

In reality, these spectra contain contributions of the stellar photosphere,
the extended atmosphere, and the circumstellar dust shell. In particular,
stellar atmospheric water ($\rm H_2O$) molecules are major absorbers in the
near-and-mid-IR range. If the stellar emission is dominated by the (cooler)
water layer ($\sim$2000 K), the temperatures estimated from spectral type are
too high. However, we have repeated our analysis using stellar temperatures in
range 2000--3500\,K (including ignoring the variation with pulsation phases)
and found that the effect on spectral feature position,  strength ratios and fitted dust temperature is
negligible.

Having subtracted the starlight from our spectra, we are left with the
emission from dust (i.e. $B_{dust}Q_{dust}$; Eq.~\ref{eqsum1}), which still
contains a temperature factor.
Therefore, we have fitted each starlight-subtracted spectrum with a blackbody
representative of the inner dust temperature ($T_{dust}$).
The estimated dust blackbody temperatures ($T_{dust}$) are listed in
Table~\ref{tab:lightcurve} and dust blackbodies are shown in Figure~\ref{fig:TCep-all7} (in dotted lines),
together with the observed spectra
and stellar continua.

\begin{figure}[t]
\begin{center}
{\includegraphics[width=0.9\textwidth]{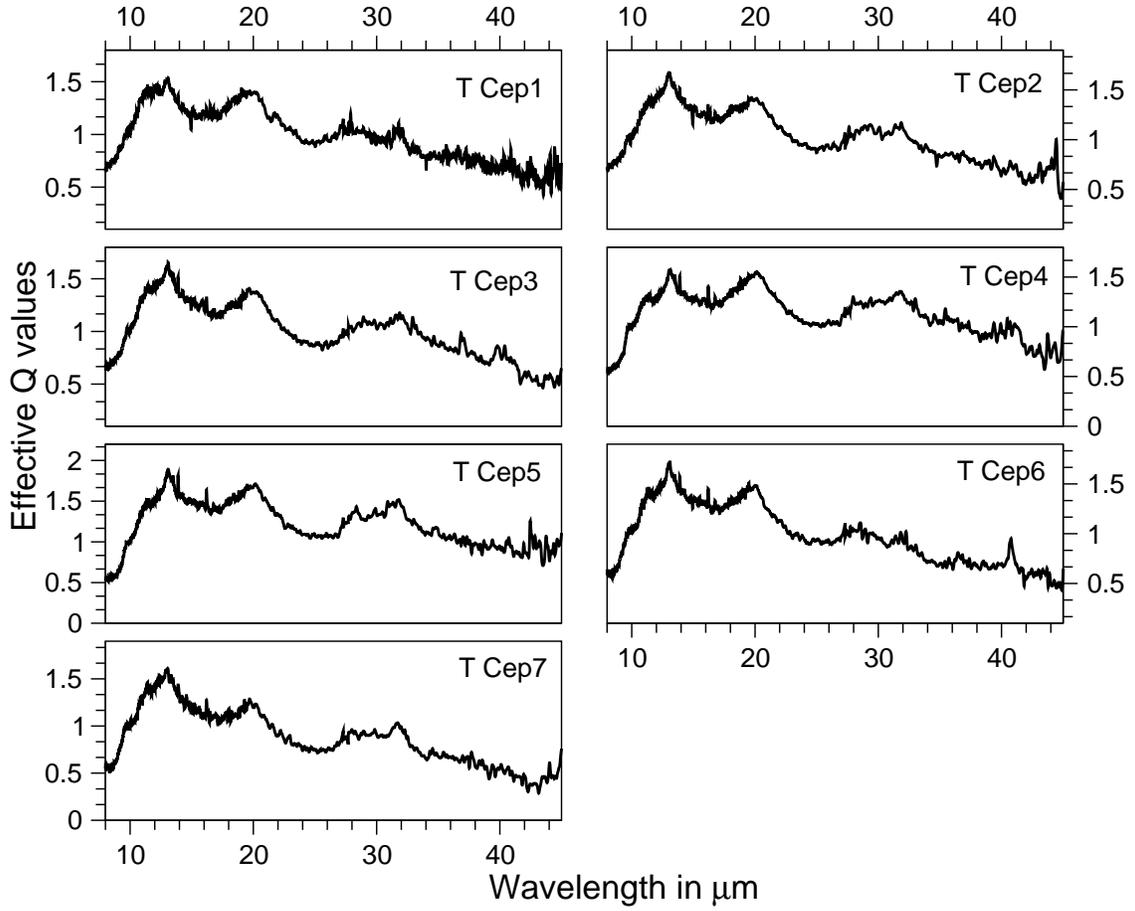}}
\caption{Starlight subtracted, dust-continuum-divided spectra (continuum-eliminated spectra).}
\label{fig:TCep_sub_Tstar_div_Tdust}
\end{center}
\end{figure}

Dividing each starlight-subtracted spectrum by the best-fitted dust
blackbody curve (normalized at 10.0\,$\mu$m, averaging over the range from 9.98--10.02\,$\mu$m)
leaves only the intrinsic
absorption/emission properties of the dust ``$Q_{dust}$'' (shown in Figure~\ref{fig:TCep_sub_Tstar_div_Tdust}),
by assuming all dust species have the same blackbody temperature.
In this case, we effectively construct a composite emission efficiency
spectrum which can be compared directly with absorptivity or mass absorption
coefficient measurements of minerals in the laboratory.
Hereafter, we refer to the starlight-subtracted, dust-continuum-divided
spectra as continuum-eliminated spectra for simplicity.

Since, T~Cep has an optically thin dust shell and its dust spectrum is dominated by the innermost dust,
we can apply this simple modeling method, rather than the more complex radiative transfer modeling method.
Radiative transfer  modeling is known to be degenerate because it actually models optical depth ($\tau$),
which convolves geometrical shell thickness, opacity and density.
Furthermore, most radiative transfer models use optical constants ($n$ \& $k$) from laboratory experiments
(or, worse, artificially derived optical constants), which are then applied usually using
Mie theory to calculate the opacities/absorption cross sections of spherical grains.
Recent studies \citep[see e.g.][]{Min-et-al-2003, DePew-et-al-2006, Pitman-et-al-2008, Corman2010}
show that the use of spherical grains leads to unrealistic spectral
features and this problem will be
discussed further in \S~\ref{grainshape}.
In addition, mineral data in the form of optical constants (complex refractive index, $n$ \& $k$)
only exist for a limited set of compositions. Available complex refractive indices are limited to the
end-members of the olivine series (forsterite, fayalite), and for
pyroxenes the situation is worse (only enstatite is available).
Consequently, using radiative transfer modeling we cannot explore the mineralogical
parameter space so thoroughly.
Although, our simple modeling approach does not account for different
sizes, shapes and/or temperatures within the dust grain population
\citep[see][for more discussion]{Thompson-et-al-2006},
it does allow us to explore the mineralogical parameter space comprehensively.

\section{Results from analysis of T~Cep's spectrum}
\label{tour}

Table~\ref{tab:lightcurve} shows that our modeled inner
dust temperature ($T_{dust}$) changes with pulsation cycle of the
star. This result is not dependent on the precise temperature of the subtracted stellar black body.
It  is unexpected because it is usually assumed that dust forms at minimum light and
the dust formation temperature is rather constant.
However, the temperature at which dust can be formed
depends on the mass-loss rate, which should vary through the stellar
pulsation cycle. Thus the changing inner dust temperature may reflect the
stability temperature at that moment in the pulsation cycle.
Consequently, our simple model suggests that dust may not be formed constantly; rather it is sporadic.
This finding is similar to that for
another M-type variable star Z Cyg \citep{Onaka-et-al-2002}.

\begin{figure}[t]
\begin{center}
{\includegraphics[width=0.7\textwidth]{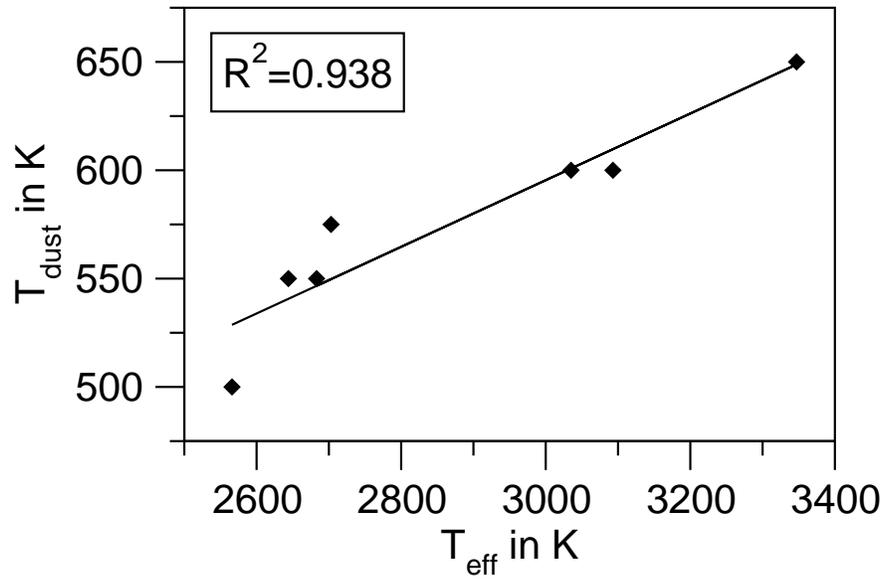}}
\caption{Correlation between assumed stellar temperature ($T_{\rm eff}$) and
fitted inner dust temperature ($T_{dust}$),
along with the linear correlation coefficient.}
\label{fig:correlation-T_eff-T_dust-V-linear}
\end{center}
\end{figure}

We have calculated the linear correlation coefficients
between the estimated stellar temperature ($T_{\rm eff}$,
 which is derived from the visual magnitude and is thus a measure of the phase of pulsation) and the fitted dust
temperature ($T_{dust}$). There is a  strong
correlation\footnote{The linear regression coefficient is $R$, but using the determination coefficient, $R^2 > 0.5$ is a better
criterion for whether a correlation exists \citep{Thompson-et-al-2006,Chan-Speck-2011}}
between them as shown in Figure~\ref{fig:correlation-T_eff-T_dust-V-linear},
suggesting that the apparent change in dust
temperature is dictated by the stellar variations, rather than the dust
formation.
This variation in inner dust temperature will be discussed further in
 \S~\ref{mixtures}.

\begin{figure}[t]
\begin{center}
{\includegraphics[width=0.5\textwidth]{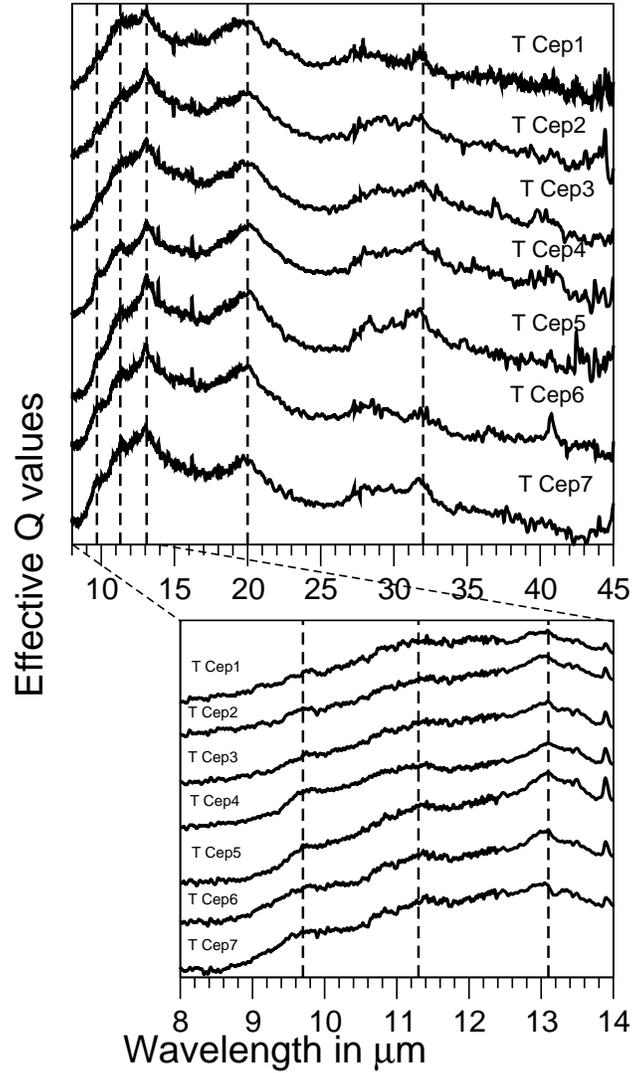}}
\caption{Continuum-eliminated spectra of T~Cep in mid-IR region.
The position of the spectral features (at 9.7, 11.3, 13.1, 20, 32\,$\mu$m) are indicated by dashed straight lines in the top panel. The sub-peak features within the broad 8--14\,$\mu$m emission (at 9.7, 11.3, 13.1\,$\mu$m) are explicitly shown in the bottom panel.}
\label{fig:TCep_sub_Tstar_div_Tdust_500K}
\end{center}
\end{figure}

In order to study the dust properties (e.g.\ grain mineralogy, morphology)
of T~Cep, we concentrate on the continuum-eliminated spectra
for the entire 8--45\,$\mu$m,
all seven of which are shown in Figure~\ref{fig:TCep_sub_Tstar_div_Tdust_500K}: top panel.
%
Since, the effect of dust temperature has been removed in our analysis
(as discussed in \S~\ref{analysis}), there is no apparent change in overall
underlying slope over the course of the pulsation period.
However, in all cases there are clear spectral features that peak at
9.7, 11.3, 13.1, 20 and 32\,$\mu$m, indicated by dashed straight lines (see
Figure~\ref{fig:TCep_sub_Tstar_div_Tdust_500K}: top panel). The sub-peak features within
the broad 8--14\,$\mu$m complex emission features (at 9.7, 11.3, 13.1\,$\mu$m) are
explicitly shown in Figure~\ref{fig:TCep_sub_Tstar_div_Tdust_500K}: bottom panel.
These sub-peak features are subtle, but we are confident that they are real because they do not occur
at any known artifact wavelength of the SWS according to the SWS handbook\footnote{%
http://iso.esac.esa.int/manuals/HANDBOOK/sws\_hb/}.

Previous studies of the \textit{ISO} SWS spectra of T~Cep discovered sharp molecular
bands at 2.5, 7.3, 16.2\,$\mu$m, which are attributed to molecular
$\rm H_2O$, $\rm SO_2$ and $\rm CO_2$ gas, respectively
\citep[e.g.][]{Yamamura-et-al-1999, Cami-et-al-1999, Matsuura-et-al-2002,
vanmalderen}. The  attribution of the 7.3\,$\mu$m feature has been further refined to be a combination of $\rm H_2O$
emission and $\rm SiO$ absorption \citep{Verhoelst-et-al-2006}.
Figure~\ref{fig:TCep-all7} shows that these molecular emission features change substantially
with the pulsation cycle of T~Cep, whereas the dust production shows only slight change.
Molecular emission is affected more strongly than the dust emission by the stellar temperature,
which changes with the pulsation cycle.
The SpectraFactory database \citep{Cami-et-al-2010} shows that molecular $\rm H_2O$ and $\rm OH$ exhibit several
spectral bands at $\lambda >20\,\mu$m. However, the molecular features are beyond the scope of the present work.

\begin{figure}[t]
\begin{center}
{\includegraphics[width=0.6\textwidth]{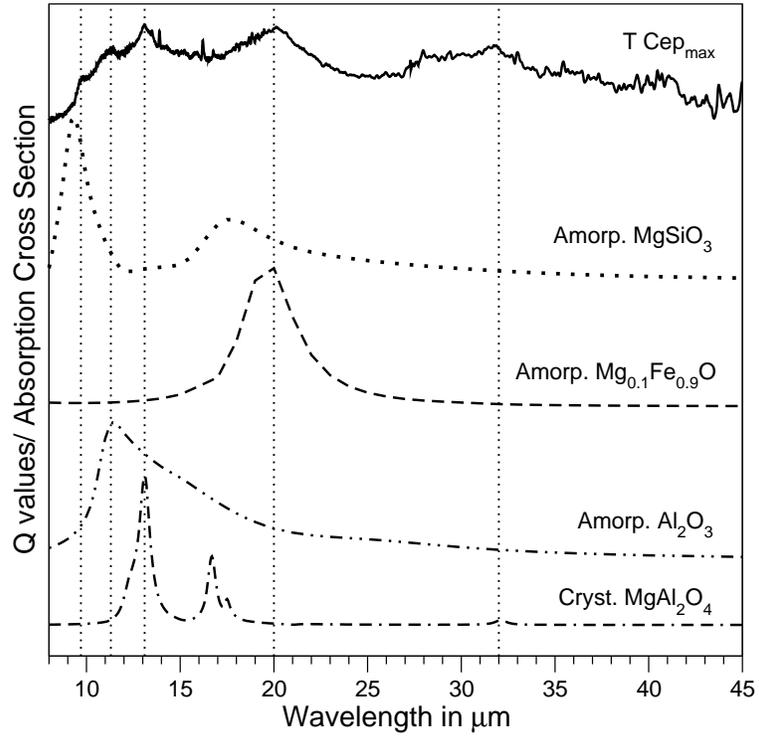}}
\caption{Comparison of continuum-eliminated spectra of T Cep4 (T~Cep$_{\rm max}$) with several amorphous minerals suggested by Van Malderen (2003). The dotted straight lines indicate the positions of dust spectral features as shown in Figure~\ref{fig:TCep_sub_Tstar_div_Tdust_500K}.}
\label{fig:comparison_van_malderen_TCep_bw}
\end{center}
\end{figure}

In addition to studying the molecular features, \citet{vanmalderen}
also studied the dust features of T~Cep, using the same \textit{ISO} SWS spectra presented
here.
He concurred that the dust spectra of T~Cep are characterized by 9.7, 11,
13 and 19.5\,$\mu$m features, but his interpretation of the carriers of these
features is different from our present analysis; he used
amorphous silicate ($\rm MgSiO_3$),
amorphous alumina ($\rm Al_2O_3$),
spinel ($\rm MgAl_2O_4$) and
amorphous magnesium-iron oxides ($\rm Mg_{0.1}Fe_{0.9}O$) respectively.
However, as discussed in \S~\ref{specclas}, spinel is no longer considered the likely
carrier of the 13\,$\mu$m feature, while $\rm Mg_{0.1}Fe_{0.9}O$
as the carrier of the 19.5\,$\mu$m feature also has some difficulties.
Furthermore, using the same optical constants
(see  Table~\ref{tab:labdata}) used by Van Malderen,
it is not possible to produce
sharp enough features to  explain those
observed in the spectrum of T~Cep
(see Figure~\ref{fig:comparison_van_malderen_TCep_bw}).

In order to understand the lattice structure and the mineralogy of the dust around
T~Cep, we have sought linear correlations amongst five prominent spectral
features at 9.7, 11.3, 13.1, 20 and 32\,$\mu$m.
To ensure that the measure of relative strength of the features is independent
of the choice of continuum fitting technique, we define the relative intensity as the ratio of the flux measured at a peak position ($F_{peak}(\lambda_1)$) to that measured at a continuum point ($F_{cont.}(\lambda_2)$).
In particular the relative intensity ($F_{peak}(\lambda_1)/F_{cont.}(\lambda_2)$) is
less likely to be affected by overlapping molecular absorption bands than any fitted continuum
\citep[especially from SiO absorption;][]{Tsuji-et-al-1997,Speck-et-al-2000}.
The peak flux intensities are measured at $\lambda_1$ = 9.7, 11.3, 13.1, 20,
32\,$\mu$m from each original flux-calibrated spectra.
We used two separate continuum points
(at $\lambda_2$ =  8.2 and 40\,$\mu$m) to verify our results. Towards the long wavelength end
of the spectra the signal-to-noise deteriorates. The increased noise could
affect our correlations. Consequently, we take the average flux in a
wavelength bin covering the range of  39.9--40.1\,$\mu$m which contains 20 points.

Table~\ref{tab:fluxratio} lists the determination coefficients ($R^2$)
for the flux ratios at peak positions ($\lambda_1$ = 9.7, 11.3, 13.1, 20, 32\,$\mu$m)
with respect to the continuum points at ($\lambda_2$ = 8.2\,$\mu$m and 40\,$\mu$m).
This table shows that not all the features have strengths which are
strongly/significantly correlated if the continuum reference is at
8.2\,$\mu$m. However, if the reference continuum point of  40\,$\mu$m is used,
almost all of the features have strongly/significantly correlated strengths.
The discrepancy between these results for the different continuum points
can be explained by the effect of SiO molecular absorption
band, which may overlap with the 8.2\,$\mu$m region and make this wavelength
point not truly continuum.
If the SiO absorption is very strong, then the FWHM of the
feature will be affected, but not necessarily the peak positions.

The strong correlations amongst all the peak features with respect to the
40\,$\mu$m continuum point (see Table~\ref{tab:fluxratio}) demonstrate that
the 32\,$\mu$m is a real feature (as also discussed in \S~\ref{specclas}),
rather than an artifact in these spectra, as
suggested by \citet{Sloan-et-al-2003}.
The strong correlation between 9.7 and 11.3\,$\mu$m features strongly suggests that the carrier is a
crystalline olivine, rather than a combination of two separate minerals (i.e.
amorphous silicate and amorphous alumina, as suggested by \citet{vanmalderen}).
Furthermore the strong correlations between the 9--12\,$\mu$m complex features and the far-IR
features at (20 and 32\,$\mu$m) provide evidence that these far-IR
features probably arise from the same carrier (i.e. crystalline olivine).
Moreover, some previous studies suggest the 13\,$\mu$m feature is carried
by a different dust species from the rest of the ``broad'' feature
(as discussed in \S~\ref{specclas}), the correlation of the 13\,$\mu$m
feature with the numerous other features supports the hypothesis that this
feature is due to some form of silicate
\citep[see][]{Begemann-et-al-1997,Speck-et-al-2000}. This will be further discussed in \S~\ref{mineralogy}.
The positions of the spectral features and correlations amongst all these features together,
suggests a crystalline silicate origin.

Having determined that the apparent broad spectral feature appears to be
composed of overlapping features at 9.7, 11.3,
13.1, 20, 32\,$\mu$m, we must now consider the implications for the mineralogy
of the dust around T~Cep.

\section{Mineralogy and morphology of dust grains around T~Cep}

\subsection{Comparing T~Cep spectrum with laboratory data of crystalline silicates}
\label{mineralogy}

As discussed above, the \textit{ISO} SWS spectra of T~Cep, a low mass-loss rate O-rich AGB star, provide
the evidence of crystalline silicates in its circumstellar
environment, which call the current dust condensation sequences into question.
From the strong correlations among the spectral features (as discussed in \S~\ref{tour})
the features at 9.7, 11.3, 20 and 32\,$\mu$m may be indicative of crystalline silicate minerals
\citep[as also mentioned by the following authors;][]{waters96, Molster-et-al-2002-a, Pitman-et-al-2010}.
While the correlations discussed in \S~\ref{tour} suggest that the carrier of the 13\,$\mu$m feature is a silicate,
we do not have an exhaustive database of silicate mineral spectra to investigate this idea.
In fact, silica-rich minerals can show a
13\,$\mu$m feature and future work to find spectra of  silica-rich
minerals may be fruitful.
Meanwhile, whereas the 20\,$\mu$m feature has been attributed to $\rm Mg_{0.1}Fe_{0.9}O$
\citep[e.g.][]{Posch-et-al-2002, Cami-2003}, our analysis suggests a silicate origin because of the correlations amongst the features.

\begin{figure}[t]
\begin{center}
{\includegraphics[width=0.6\textwidth]{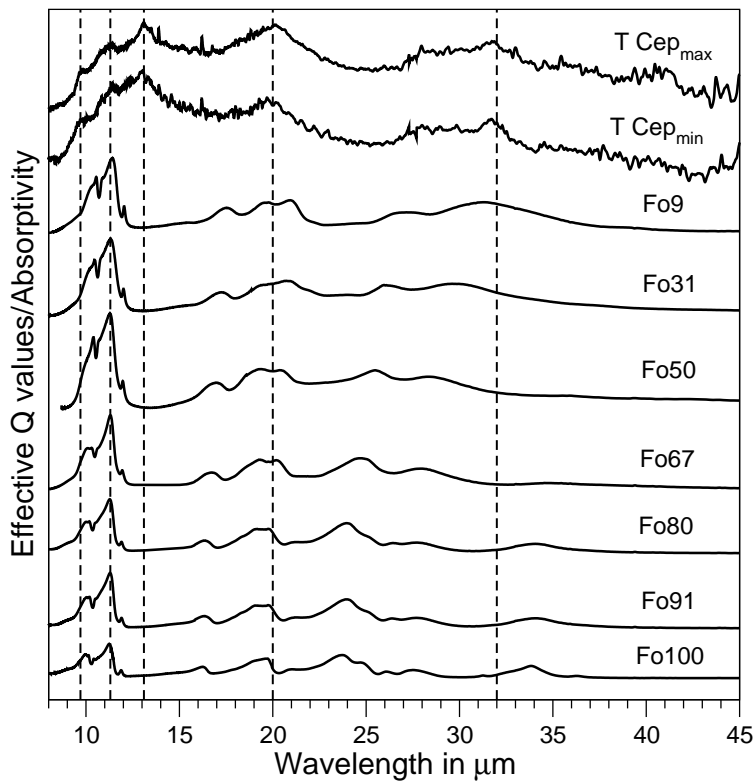}}
\caption{Comparison of continuum-eliminated spectra at maximum (T~Cep$_{\rm max}$) and minimum (T~Cep$_{\rm min}$) light with the laboratory absorptivity data for
the olivine solid solution with varying Fe/(Mg+Fe) ratios.
FoX (X=9, 31, 50, 67, 80, 91, 100) defines the composition such that
each olivine has the composition Mg$_{2X/100}$Fe$_{2-2(X/100)}$SiO$_4$.
The dotted straight lines indicate the positions of dust spectral features as shown in Figure~\ref{fig:TCep_sub_Tstar_div_Tdust_500K}.}
\label{fig:comparison-TCep4-TCep7-different-olivine-modified-bw}
\end{center}
\end{figure}

\begin{figure}[t]
\begin{center}
{\includegraphics[width=0.6\textwidth]{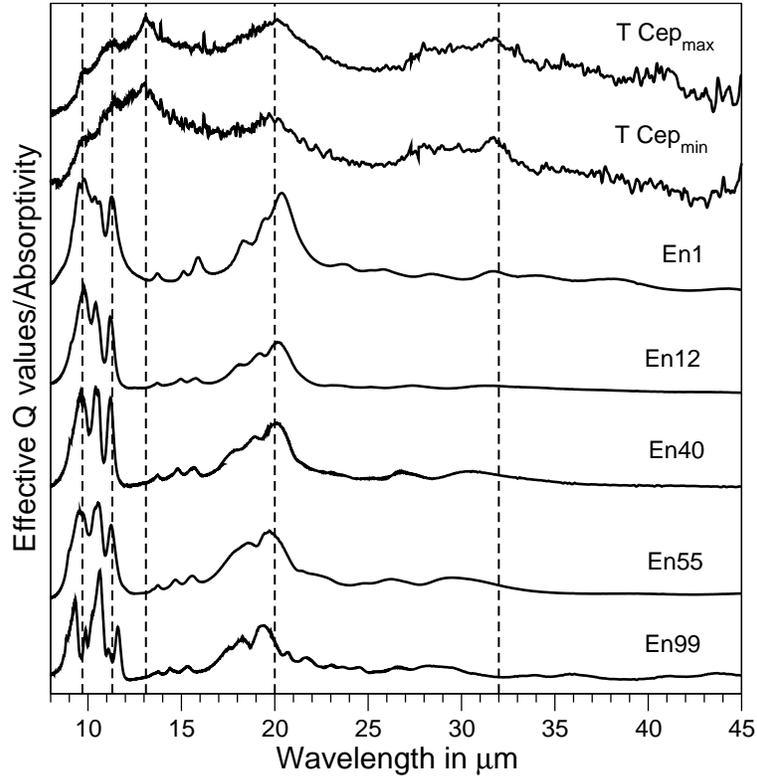}}
\caption{Comparison of continuum-eliminated spectra at maximum (T~Cep$_{\rm max}$) and minimum (T~Cep$_{\rm min}$) light with the laboratory absorptivity data for
a pyroxene solid solution with varying Fe/(Mg+Fe) ratios.
EnX (X=1, 12, 40, 55, 99) defines the composition such that
that each pyroxene has the composition Mg$_{X/100}$Fe$_{1-(X/100)}$SiO$_3$.
The dotted straight lines indicate the positions of dust spectra features as shown in Figure~\ref{fig:TCep_sub_Tstar_div_Tdust_500K}.}
\label{fig:pyroxene_hofmeister_TCep4_TCep7}
\end{center}
\end{figure}

In order to match and identify the dust species present in the circumstellar envelope of T~Cep,
we compare the spectral features of the laboratory data from
\citet{Pitman-et-al-2010} and \citet{Hofmeister-et-al-2007} (hereafter WashU Group)
of different crystalline dust species to the \textit{ISO} spectra of T~Cep.
In Figure~\ref{fig:comparison-TCep4-TCep7-different-olivine-modified-bw}, we
compare the continuum-eliminated spectra of
T~Cep$_{\rm max}$ and T~Cep$_{\rm min}$ with the laboratory absorptivity data for a selection of
crystalline olivine samples
of varying composition (data taken from \citet{Pitman-et-al-2010}).
Likewise, Figure~\ref{fig:pyroxene_hofmeister_TCep4_TCep7} compares
the same two T~Cep spectra together with laboratory absorptivity data
for a selection of crystalline pyroxene
samples of varying composition (data taken from Hofmeister et al. in prep.)\footnote{%
We have a longstanding collaboration with Anne Hofmeister at WashU and can
access new laboratory data through:
http://galena.wustl.edu/$\sim$dustspec/info.html, even before publication.
These crystalline olivine and pyroxene series spectra  agree well with the laboratory data from Kyoto group \citep[e.g.][]{Koike-et-al-2003, Chihara-et-al-2002} but provide fine grid spacing in composition space. Comparison among the data from several laboratory groups will be further discussed in \S~\ref{grainshape}.}

For the olivine data, Fo$X$ is an
indication of the composition such that each olivine has the composition
Mg$_{2X/100}$Fe$_{2-2(X/100)}$SiO$_4$ (dataset ranging from Fo9 to Fo100 are shown in Figure~\ref{fig:comparison-TCep4-TCep7-different-olivine-modified-bw});
for the pyroxene data, En$X$ gives
the composition via Mg$_{X/100}$Fe$_{1-(X/100)}$SiO$_3$ (dataset ranging from En1 to En99 are shown in Figure~\ref{fig:pyroxene_hofmeister_TCep4_TCep7}).
Since we have compiled laboratory mineral spectral data from a number of sources, these sources,
along with other relevant sample/experimental information are listed in
Table~\ref{tab:labdata}.
The WashU laboratory spectra are in the form of
absorbance ($a$), which is proportional to the optical depth,
whereas we need to
compare to the absorption efficiency ($Q_{\rm abs}$). In order to convert the
laboratory data to an appropriate form we use absorbance ($a$) is proportional
to the log of the absorptivity ($A$), i.e. $a \propto {\rm ln}(A)$.

In general, we compare the spectral features of the laboratory spectra of different dust
species to the astronomical observational data in order to match and identify the dust
species present in circumstellar envelopes. However, the spectral feature parameters
(positions, strengths and widths) are significantly influenced by three
parameters (composition, temperature and grain shape of the dust grains).
The effect of these three parameters are discussed below.

It is clear from both Figures
\ref{fig:comparison-TCep4-TCep7-different-olivine-modified-bw}
(for the olivine family members) and
\ref{fig:pyroxene_hofmeister_TCep4_TCep7}
(for the pyroxene family members) that positions and strengths of the spectral features
change with varying Fe/[Mg+Fe] ratio. The positions of peak features shift
towards longer wavelength as the Fe/[Mg+Fe] ratio increases. For
T~Cep, the spectral features are more closely matched to Fe-rich silicate dust (Fo9 and En1),
rather than expected Mg-rich silicate dust (Fo100 or En99; see Figure~\ref{fig:comparison-TCep4-TCep7-different-olivine-modified-bw}
and \ref{fig:pyroxene_hofmeister_TCep4_TCep7}), which calls the
conventional wisdom regarding the dust condensation sequence
into question.
These laboratory data compared to the observational spectra of T~Cep, preclude the
possibility of large abundances of the conventional Mg-rich silicates
(Fo100, En99) to explain the dust features, unless
grain shape or temperature effects can be invoked.

\citet{Koike-et-al-2006} performed laboratory experiments to determine the
effect of temperature on crystalline olivines.
In general increasing temperature moves spectral features redwards as well as
broadening and diminishing the heights of the features.
This effect is most marked for far-IR features (at 49 and 69\,$\mu$m for
forsterite), and is much less effective in the mid-IR (8-45\,$\mu$m) region.
The experiments covered a wide range of sub-room temperatures (8--292\,K), and even this
large change in temperature has little effect on the mid-IR features.
Consequently, extrapolation of the effects to higher temperature still suggests that temperature is an
insignificant factor in determining the positions of the spectral features with which we are concerned.

The third parameter often invoked to shift spectral features is grain shape
\citep[see e.g.][]{Fabian-et-al-2001, DePew-et-al-2006, Sloan-et-al-2006}.
This will be discussed in detail in \S~\ref{grainshape}.

\subsection{Investigation of mineralogy of dust around T~Cep: Compositional
mixtures}
\label{mixtures}

It is clear from the qualitative comparisons in
Figures~\ref{fig:comparison-TCep4-TCep7-different-olivine-modified-bw} and
\ref{fig:pyroxene_hofmeister_TCep4_TCep7} that Fe-rich
crystalline silicates (Fo9, En1) are promising constituents for the dust around T~Cep.

\begin{figure}[t]
\begin{center}
{\includegraphics[width=0.6\textwidth]{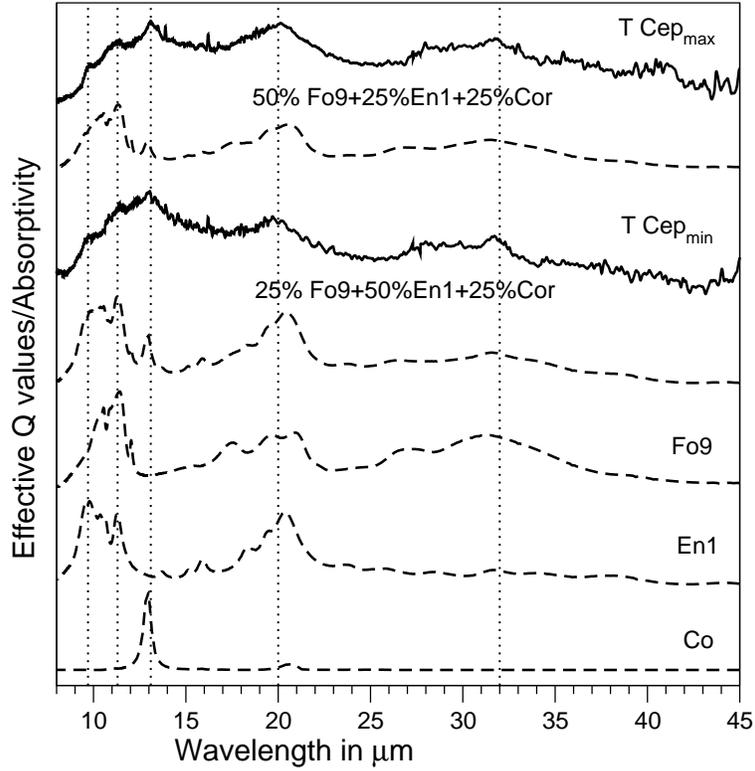}}
\caption{Comparisons of continuum-eliminated spectra at maximum (T~Cep$_{\rm max}$) and minimum (T~Cep$_{\rm min}$) light (solid lines) along with the best fit model (dashed lines) of mixtures of different potential crystalline minerals with different ratios. The laboratory data of individual crystalline minerals (Fo9, En1, Cor) are also included.
The dotted straight lines indicate the positions of dust spectra features as shown in Figure~\ref{fig:TCep_sub_Tstar_div_Tdust_500K}.}
\label{fig:comparison_mixture_TCep}
\end{center}
\end{figure}

Figure~\ref{fig:comparison_mixture_TCep} shows comparison of
continuum-eliminated spectra of T~Cep$_{\rm max}$ and
T~Cep$_{\rm min}$ with mixture of these potential
Fe-rich crystalline silicates minerals in different ratios.
To account for the 13\,$\mu$m feature, we have also included spherical
corundum (Al$_2$O$_3$) grains. We use this mineral simply to give a feature,
and not because we necessarily believe the carrier to be corundum.
The source of the laboratory data of corundum (Cor) and its chemical
composition are listed in Table~\ref{tab:labdata}.

The best fit models from our calculations are mixtures of crystalline fayalite
(Fo9; not-quite-endmember of olivine family), ferrosilite (En1) and corundum
(Cor) in varying proportion (with the error-bars of 10\% for each constituent).
Although the models are not a perfect fit, there appears to be a variation from T~Cep$_{\rm max}$ to
T~Cep$_{\rm min}$. This variation is best explained by changing the ratio of Fo9 to En1.
This is demonstrated in Figure~\ref{fig:comparison_mixture_TCep}, which shows the best fit models with
T~Cep$_{\rm max}$ and T~Cep$_{\rm min}$ along with the laboratory data of each constituent separately.

Because laboratory optical properties of solids are produced for a variety
of applications, the various laboratory data are published in several different
types of units (e.g. absorbance, mass absorption coefficient, complex refractive index etc.)
Using the appropriate conversion factors, all are converted to absorptivity ($A$) before
attempting to make mixtures for spectral fitting.

For Fo9 and En1 the original spectra were in absorbance ($a$) units
\citep[e.g. Hofmeister et al. in prep;][]{Pitman-et-al-2010}, which were
converted to absorptivity ($A$), using $A \propto e^{a}$. And for Cor, we used the \citet{Min-et-al-2003} method to
calculate the absorption cross section ($C_{\rm abs}$) for spherical grains of size 0.1\,$\mu$m using its optical constants  \citep[data taken from][]{Gervais-1991}.
Absorptivity ($A$) is directly proportional to the absorption efficiency
$Q_{\rm abs}$ of a grain, which, in turn, is directly proportional to the
absorption cross section ($C_{\rm abs}$).

We use the following linear conversion equation to convert $C_{\rm abs}$ to $A$
for corundum:

\begin{equation}
\label{eqs4}
A = C_{abs} \times n \times d
\end{equation}

\noindent
where $n$ is the number of particle per unit volume of the material used, and $d$ is the path length. For thin film, $d$ is taken as 1 $\mu$m. And to calculate $n$, we use

\begin{equation}
\label{eqs5}
n = \frac {\rho} {M_{mol} \times m_{H_{2}}}
\end{equation}

\noindent
where $\rho$ is the density of corundum (4.02 gm/cm$^3$), $m_{H_2}$ is
1.672$\times$10$^{-24}$ gm. And $M_{mol}$ for Al$_2$O$_3$ is
(27$\times$2+16$\times$3=102). Using these values in Eq~\ref{eqs5}, we
get $n$ = 2.35$\times$10$^{22}$ cm$^{-3}$. Knowing  $C_{abs}$, $n$ and $d$,
we calculated $A$ for Al$_2$O$_3$ by using Eq~\ref{eqs4}.

Figure~\ref{fig:comparison_mixture_TCep} demonstrates that these mixtures
produce a reasonably good match to the detailed shape and as well as overall
shape of the spectrum for both T~Cep$_{\rm max}$ and T~Cep$_{\rm min}$.
Consequently, we can conclude that
the compositions of the individual dust constituents remain the same in both
spectra, while the relative amounts of these constituents may change, but are consistent with no variations at all.
The spectra are consistent with a variation in the  ratio of Fo9/En1 such that it is doubled at maximum light
(T~Cep$_{\rm max}$) and the
ratio is halved at minimum light (T~Cep$_{\rm min}$), while the relative abundance of Cor remains unchanged.
This suggests  that the olivine grains are slightly favored at maximum light.


 The apparent subtle changes in mineralogy with pulsation cycle may be entirely due to
statistical effects. However, in the classic condensation sequence
(shown in Figure~\ref{fig:CondSeq2}), Mg-rich olivine exists at higher temperatures than
Mg-rich pyroxene. Furthermore the Mg-rich olivine reacts with SiO gas to form
Mg-rich pyroxene. If a similar reactive process exists for the iron-rich
endmembers, the occurrence of higher olivine/pyroxene compositions at maximum light, may be a temperature effect.
In this case, the Fe-rich pyroxene (En1) is more
easily destroyed than the Fe-rich olivine (Fo9) at maximum light;
while at minimum light the reaction with SiO gas is promoted.
Essentially, the change in stellar
radiation field causes selective destruction/processing of the inner most
grains.

From our analysis, the most striking result is that we need Fe-rich crystalline silicates in order
to explain the spectral features. This result is unexpected, but supported by recent
studies of silicate presolar grains from meteorites which show that silicate
grains from AGB stars have nano-crystalline structures and some are very
iron-rich, indicative of non-equilibrium formation processes
\citep[e.g.][]{stroud08,vollmer08,bose10}.

\subsection{Investigation of morphology of dust around T~Cep: Grain shape effects}
\label{grainshape}

Our analysis on dust compositions around T~Cep strongly suggest that the
crystalline silicates are almost completely Fe-rich, with little evidence of
Mg-rich silicates.
This result conflicts with both current dust formation hypotheses, and
studies of cosmic crystalline silicates to date.
However, as mentioned in \S~\ref{mineralogy}, it is possible that grain shape
effects may allow Mg-rich silicates to match
the positions and strengths of the observed dust features of T~Cep.
Previous studies show that the position of spectral features of dust
grains depend on their grain shapes
\citep[see][]{bh83, Bohren-et-al-1983, Fabian-et-al-2001, Min-et-al-2003, Sloan-et-al-2006, DePew-et-al-2006, Koike-et-al-2010}.
Here we investigate the grain shape effects of crystalline silicates.

The spectral features of crystalline silicate grains exhibit wide variety
in the positions, widths and strengths of their peaks.
For a given series, these are dependent on the composition (Fe/[Mg+Fe] ratio) of the silicates,
as well as the temperatures (as discussed in \S~\ref{mineralogy}) and shapes of dust grains.
Disentangling these competing effects is difficult
because we do not have the data to test both grain shape and compositional
effects simultaneously. There are several laboratories that have produced
spectral data for crystalline silicates in the olivine and pyroxene series.
The three laboratories, which are most noted for these studies are
WashU \citep[e.g.][]{Hofmeister-et-al-2007,Pitman-et-al-2010},
Kyoto \citep[e.g.][]{Chihara-et-al-2002,Koike-et-al-2003,Koike-et-al-2006,Murata-et-al-2009-b,Koike-et-al-2010} and
Jena \citep[e.g.][]{Fabian-et-al-2001,Jaeger-et-al-1998} groups.

A series of opacity (mass absorption coefficients) measurements for the olivine series, covering several
Fe/[Mg+Fe] ratios from forsterite (Fo100) to fayalite (Fo0) has been published by the Kyoto group \citep{Koike-et-al-2003}.
A finer grid of compositions across the olivine series was published by the
WashU group \citep{Hofmeister-et-al-2007,Pitman-et-al-2010}, which agree
with the Kyoto data. Here we use to finer grid of
\citet{Pitman-et-al-2010}.
A similar series of measurements for
pyroxene from enstatite (En100) to ferrosilite (En1) has been also published by the Kyoto group \citep{Chihara-et-al-2002}. Here, we use currently unpublished pyroxene data
from the WashU group (Hofmeister et al. in prep.),
which also agree with the Kyoto data.

While the data from WashU and Kyoto groups provide a good sampling over several
Fe/[Mg+Fe] ratios for both olivines and pyroxenes, they do not measure the
complex indices of refraction or dielectric constants for all these
compositions. As a consequence we cannot use them to analyze the effect of
grain shape on dust spectra. Thus we cannot fit both the
compositional effect (Fe/[Mg+Fe] ratio) and the shape distribution simultaneously.
For this reason, extensive studies of grain shape effects have generally been
limited to a single composition: forsterite (Fo100), supplied by the Jena group.

The complex refractive indices for the end-member compositions of olivine (Mg-rich olivine: Fo90\footnote{%
Technically Fo90 is still called forsterite and most natural samples are
closer to Fo90 than Fo100}, fayalite: Fo0)
and Mg-rich pyroxene (enstatite: En100) are provided by the Jena group\footnote{optical constants available at http://www.astro.uni-jena-de/Laboratory/Database/databases.html} \citep[e.g.][]{Fabian-et-al-2001, Jaeger-et-al-1998}.
The refractive indices are provided for the vibrational directions parallel to
the three crystallographic axes $x$, $y$, $z$.
Meanwhile, \citet{Mukai-et-al-1990} provided optical constants for forsterite (Fo100),
but this data was for unoriented samples and therefore
represents an average of the three axial directions.
A list of all the sources of these laboratory data used,
can be found in Table~\ref{tab:labdata}.

Since solid particles in astrophysical environments are expected to be very
irregular in shape, it is very difficult to characterize the shape of the
particles in a simple way. Traditionally, cosmic dust grains have been
assumed to be spherical for simplicity, however, it is becoming increasingly
obvious that the use of spherical grains leads to unrealistic spectral features
\citep[see e.g.][]{Min-et-al-2003, DePew-et-al-2006, Pitman-et-al-2008, Corman2010}.
There are several possible approaches to addressing the grain shape effect, which we
will compare here.

\citet{Fabian-et-al-2001} used their optical constants of crystalline Mg-rich
olivine (Fo90) and fayalite (Fo0) to calculate mass absorption coefficients for IR region
for a variety of grain shapes. These authors primarily chose: spherical grains,
ellipsoids elongated along the $z$-axis, and distributions of
ellipsoids; based on the assumption that mineral forms crystalline
ellipsoids elongated along the $z$-axis, when condensing
from the gas phase at high temperature.
Whether there should be a preferred
growth axis for a grain shape or not, will be discussed below. These authors compared their
calculated mass absorption coefficients data directly with \textit{ISO} SWS observational data of evolved stars.
They concluded that the spectral features are shifted in wavelength by grain shape effects and this effect is prominent in mid-IR region, whereas, the spectral features in far-IR region remain practically unaffected by the particle
shape.

Recent experimental studies by \citet{Takigawa-et-al-2009} on the evaporation of single crystals of
forsterite (Fo100) show that this mineral evaporates anisotropically,
which may lead to distinct grain shape
distributions (e.g. disk-shaped dust grain, flattened along the $y$ or $z$ axes).
They showed that forsterite always evaporates
anisotropically, but the details of
the anisotropy (i.e. which axis is elongated) depends on the experimental conditions
(total gas pressure and temperature).
They conclude that the precise shape of forsterite grains depends on the formation conditions in space.
Because the  peak positions and relative strengths of dust spectral features are expected to depend on grain shape, these spectral parameters may be diagnostic of the formation and  heating conditions for the dust grains.

Another recent study by the Kyoto group investigated how the spectral features of forsterite (Fo100) are affected by  grain shape
\citep{Koike-et-al-2010}. They measured the IR
mass absorption coefficients of forsterite grains of a variety of shapes including irregular, plate-like with no sharp edges, elliptical, cauliflower and spherical.
They concluded that spectral features of forsterite in mid-IR region (at 11, 19, 23, 33\,$\mu$m) are extremely sensitive to particle shape, whereas, the features in far-IR region (at 49, 69\,$\mu$m) remained unchanged despite of the different grain shapes \citep[c.f.][]{Fabian-et-al-2001}.

\citet{Min-et-al-2003} adopted statistical approach \citep[following][]{bh83} in order
to understand the effect of grain shape on IR dust
features of forsterite (Fo100). They considered the limiting case,
where the particles are very small compared to the
wavelength of radiation (Rayleigh domain).
In this statistical approach, the scattering and absorption properties of irregularly
shaped particles can be simulated by the average properties of a distribution
of simple shapes (such as ellipsoids, spheroids, and hollow spheres). They calculated
the absorption and scattering cross sections of different grain shapes and concluded
that there is a strong similarity between the absorption spectra of
distributions of various non-spherical homogeneous particles (e.g. ellipsoid, spheroids) and
a distribution of hollow spheres in Rayleigh domain, but
that homogenous spherical particles show a markedly different spectrum.

\begin{figure}[t]
\begin{center}
{\includegraphics[width=0.95\textwidth]{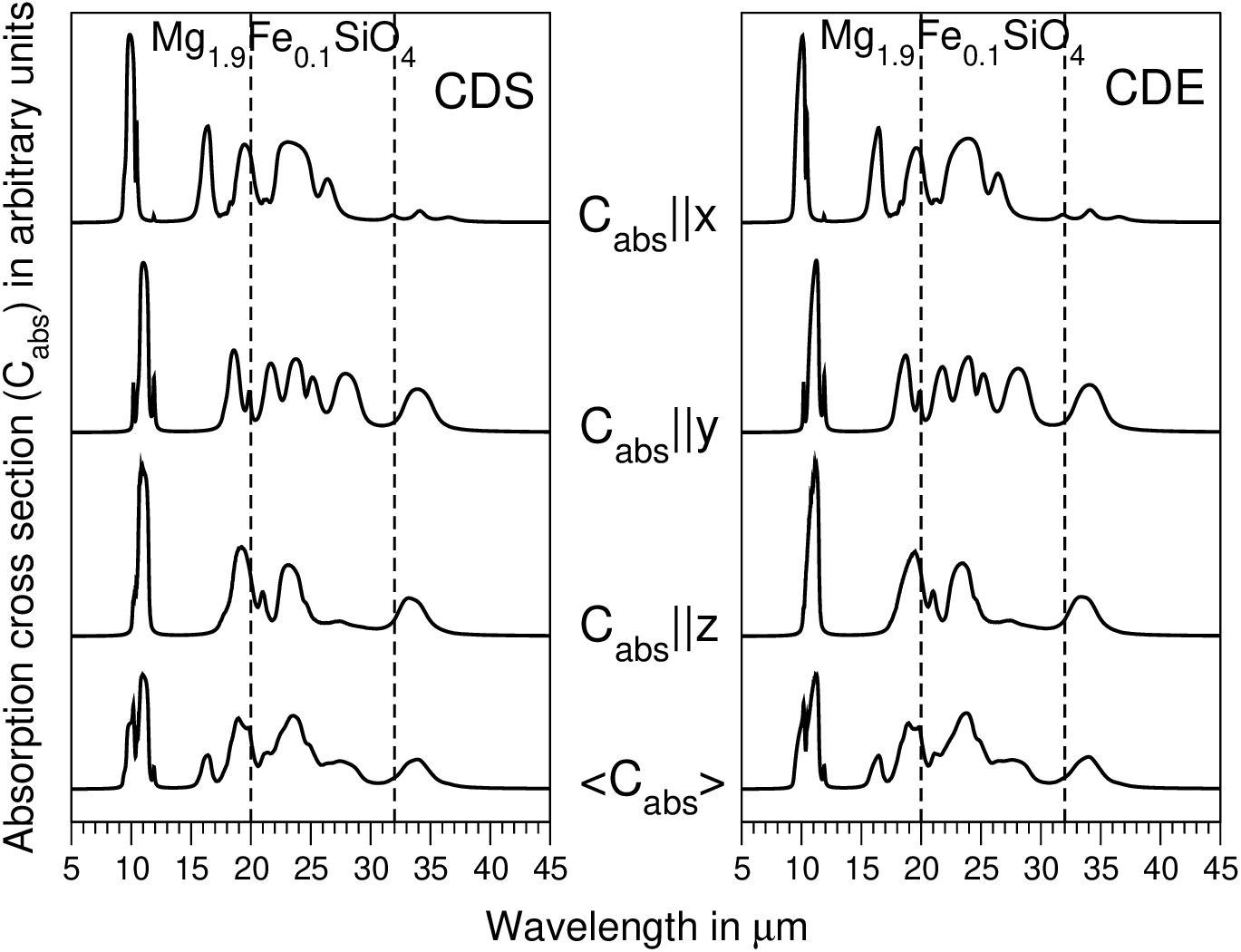}}
\caption{Absorption cross section of very Mg-rich Olivine ($\rm Mg_{1.9}Fe_{0.1}SiO_4$) for
spheroid (CDS: left panel) and ellipsoid (CDE: right panel). The upper three plots correspond to calculated $C_{\rm abs}$ parallel to the $x$, $y$ and $z$ axes. The bottom plots correspond to average ($\langle C_{\rm abs} \rangle$).
Vertical dashed lines show positions of observed spectral features of T~Cep at long wavelength (20 and
32\,$\mu$m).}
\label{fig:C_abs-Mg-rich-olivine-CDS-CDE}
\end{center}
\end{figure}

\begin{figure}[t]
\begin{center}
{\includegraphics[width=0.95\textwidth]{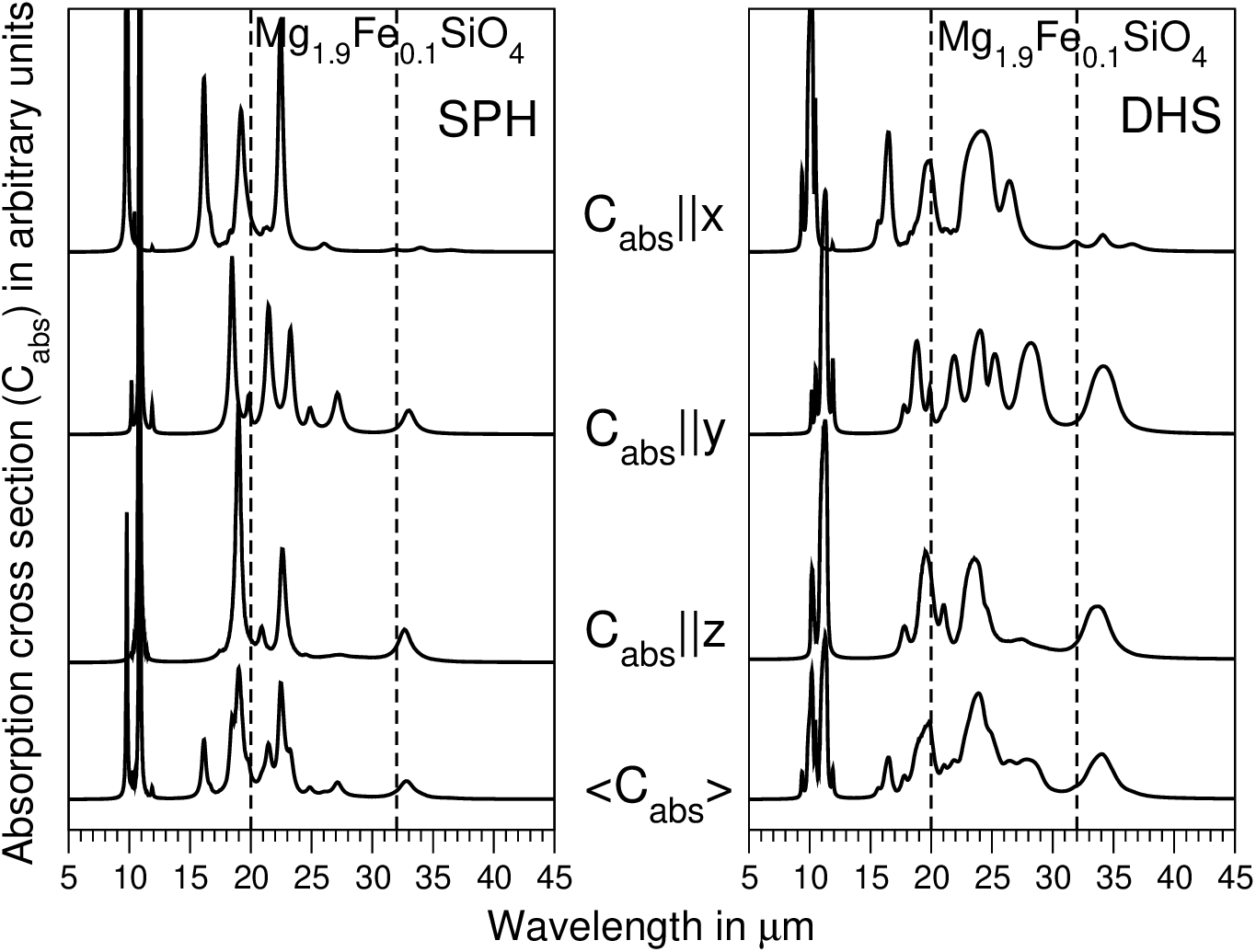}}
\caption{Absorption cross section of very Mg-rich Olivine ($\rm Mg_{1.9}Fe_{0.1}SiO_4$) for
spherical (SPH: left panel) and hollow sphere (DHS: right panel). The upper three plots correspond to calculated
$C_{\rm abs}$ parallel to the  $x$, $y$ and $z$ axes. The bottom plots correspond to average ($\langle C_{\rm abs} \rangle$).
Vertical dashed lines show positions of observed spectral features of T~Cep at long wavelength (20 and
32\,$\mu$m).}
\label{fig:C_abs-Mg-rich-olivine-SPH-DHS}
\end{center}
\end{figure}

\begin{figure}[t]
\begin{center}
{\includegraphics[width=0.95\textwidth]{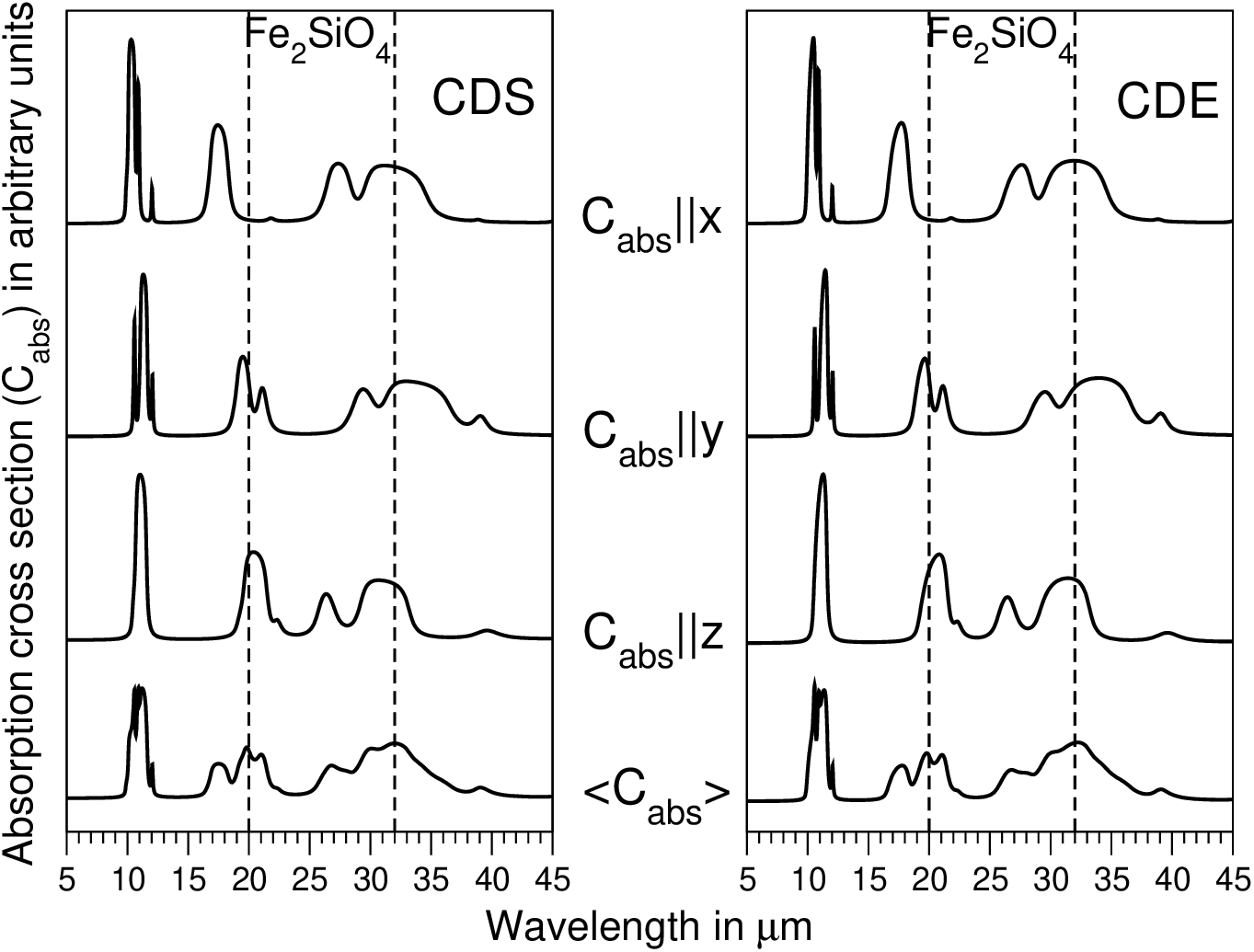}}
\caption{Plot of Absorption cross section of fayalite ($\rm Fe_2SiO_4$) for spheroid (CDS: left panel) and ellipsoid (CDE: right panel). The upper three plots correspond to calculated $C_{\rm abs}$ parallel to the $x$, $y$ and $z$ axes. The bottom plots correspond to average ($\langle C_{\rm abs} \rangle$) of them.
Vertical dashed lines show positions of observed spectral features of T~Cep at long wavelength (20 and
32\,$\mu$m).}
\label{fig:C_abs-fayalite-CDS-CDE}
\end{center}
\end{figure}

\begin{figure}[t]
\begin{center}
{\includegraphics[width=0.9\textwidth]{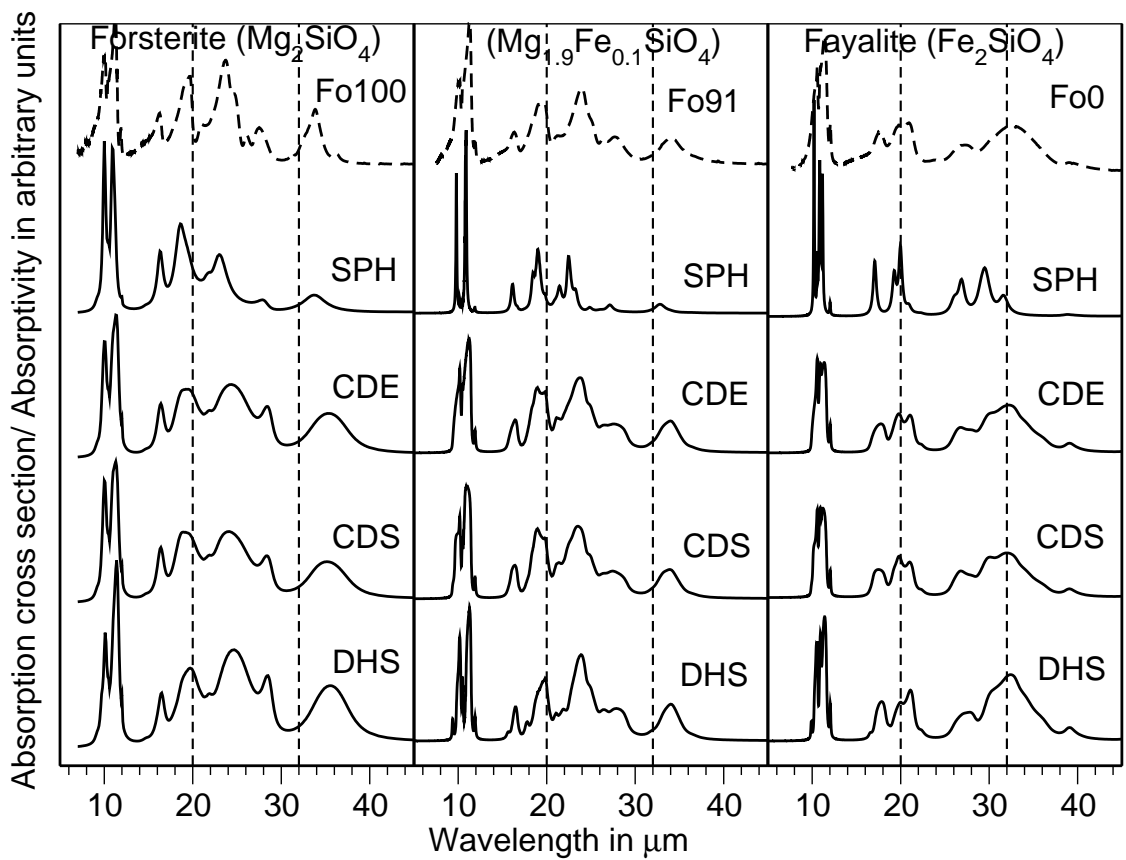}}
\caption{Calculated absorption cross section ($C_{\rm abs}$) for four shape
distributions (SPH, CDE, CDS, DHS) of 0.1\,$\mu$m-sized grains of three different members of olivine family along with the directly-measured laboratory absorption spectrum.
{\it Right panel}: forsterite (Fo100),
{\it center panel}: very Mg-rich
olivine (Fo91: closest compositionally in the lab spectra),
{\it left panel}: fayalite (Fo0).
Vertical dashed lines show positions of observed spectral features of T~Cep at long wavelength (20 and
32\,$\mu$m).
The solid lines are the calculated absorption cross-sections while
dashed lines (at the top) show the laboratory absorption spectra
\citep[data taken from][]{Pitman-et-al-2010}.}
\label{fig:comparison-grainshape-olivine-hofmeister}
\end{center}
\end{figure}

Using the approach of \citet{Min-et-al-2003} and assuming that there
is no tendency for a certain axis to be elongated, we have calculated the absorption
cross-sections ($C_{\rm abs}$) for three compositions of olivine (Fo100, Fo90 and Fo0; see Table~\ref{tab:labdata}
for sources of complex refractive indices) .
For each composition, we calculate $C_{\rm abs}$ for four grain shape distributions:
Spherical particle (SPH),
Continuous distribution of ellipsoids (CDE),
Continuous distribution of spheroids (CDS), and
Distribution of hollow spheres (DHS)\footnote{%
Hollow spheres are meant to simulate fluffy grains}.
The grain size is defined by the grain volume being equivalent to a sphere of radius 0.1\,$\mu$m.
When data is available for the individual crystallographic axes, we calculate the  $C_{\rm abs}$  separately and then
average the three axial absorption cross sections ($\langle C_{\rm abs} \rangle$).
The final $\langle C_{\rm abs} \rangle$ spectra are shown in Figures~\ref{fig:C_abs-Mg-rich-olivine-CDS-CDE}--\ref{fig:comparison-grainshape-olivine-hofmeister}.

For all three compositions (Fo100, Fo90 and Fo0), the spectral feature parameters
(e.g. peak positions and strengths) are similar for both ellipsoids (CDE) and spheroids (CDS).
For the hollow spheres (DHS) the peak positions are similar to both CDE and CDS, but the spectral features
are stronger. The spectral features of spherical (SPH) particles
are significantly different in both position and strength from those of CDE, CDS and DHS.

Figure~\ref{fig:comparison-grainshape-olivine-hofmeister} compares the positions of the spectral features in the derived absorption cross-sections with those observed for T~Cep. It is clear that fayalite (Fo0) provides  a better
match with the observed peak positions (at 20, 32\,$\mu$m) of T~Cep, than does forsterite (Fo100).
Following the work of \citet{Takigawa-et-al-2009}, it is possible that a single growth axis is elongated. However, Figure~\ref{fig:comparison-grainshape-olivine-hofmeister} shows that even using only a single crystallographic axis does not produce an Mg-rich olivine with features that match T~Cep.

\section{Discussion}

Our analysis of the spectra of T~Cep strongly suggest that the dust forming around this
Mira is both highly crystalline and iron-rich.
This has implications for not only for other stars exhibiting these crystalline features,
but also for conventionally Mg-rich condensation sequence observed for amorphous circumstellar silicates.

\citet{woitke-06} showed that
while carbon dust around an AGB star could drive a radiation-pressured
wind, oxygen-rich dust (silicate) was too transparent.
However, this result assumes the silicates are Mg-rich.
Iron-rich silicate grains tend to have higher optical/near-IR opacities which
facilitate the capture of momentum from the star through radiation pressure.
The inclusion of iron-rich silicate grains may solve this problem.
%

\citet{woitke06} also showed that
the dynamics in the dust-forming zones around
carbon-rich AGB stars lead to inhomogeneous dust formation, producing fine scale
structure in the density of the dust envelope. In these models the only
condensate considered is amorphous carbon. In an oxygen-rich environment, there
are many potential minerals that can be formed and their stability is sensitive to
the precise conditions.
In addition to the turbulent/hydrodynamic density inhomogeneities predicted by \citet{woitke-06},  pulsation shocks are expected to have a strong effects
on local conditions \citep[e.g.][]{cherchneff06}. This combination of physical effects should lead to non-equilibrium dust formation and may lead to unexpected dust-forming conditions.
Therefore we suggest that even at low mass-loss rates the density structure in the outflows of AGB
stars is such that crystalline silicates may form, though it is not clear why
iron-rich silicates are apparently favored.

\section{Summary \& Conclusion}

We have presented an analysis of the time variations of the IR dust spectrum of
optically thin O-rich AGB star, T~Cep.

We found that:
\begin{enumerate}

\item{
The inner dust temperature of T~Cep is variable.}

\item{
Dust formation is likely to be sporadic, not continuous,
and has approximately the same composition all the time.}

\item{
While the observations are consistent with a constant  olivine-to-pyroxene ratio, they can accommodate small variations with
stellar pulsation.
This variation in composition is very subtle
and can be explained by selective destruction/processing of the inner most grains with the
change in stellar radiation field.}

\item{
The strong correlations between the observed spectral features suggest
that they all have same crystalline mineral as a carrier.}

\item{
The structure within the broad 8--14\,$\mu$m feature with overlapping
sub-features at 9.7, 11.3, 13.1\,$\mu$m is explained
by mixtures of crystalline silicates. This confirms the
presence of crystalline minerals around low-mass-loss rate O-rich AGB stars.}

\item{
The peak wavelength of the features at 20 and 32\,$\mu$m  suggest
the presence of Fe-rich, rather than the expected Mg-rich silicates. This can
be explained as occurrence of non-equilibrium condensation mechanism in the
outflow of the central star.}

\end{enumerate}

The analysis presented here shows that our understanding of the formation of
crystalline silicates and the inclusion of iron in those silicates is in its
infancy and needs to be revised according to the present findings.

Acknowledgements:
We are grateful to the anonymous reviewers, whose comments significantly improved this paper. And we thank Dr. Catharinus Dijkstra for providing his code to calculate the absorption cross section for four different grain shapes of various minerals.

This work is supported by NSF CAREER AST-0642991.




\clearpage

\clearpage

\clearpage

\clearpage

\clearpage

\clearpage

\clearpage

\clearpage

\clearpage

\clearpage

\clearpage

\clearpage

\clearpage

\clearpage
\begin{figure}[t]
\begin{center}
{\includegraphics[width=0.95\textwidth]{fig14.eps}}
\caption{Plot of Absorption cross section of fayalite ($\rm Fe_2SiO_4$) for spherical (SPH: left panel) and hollow sphere (DHS: right panel). The upper three plots correspond to calculated $C_{\rm abs}$ parallel to the $x$, $y$ and $z$ axes. The bottom plots correspond to average ($\langle C_{\rm abs} \rangle$).
Vertical dashed lines show positions of observed spectral features of T~Cep at long wavelength (20 and
32\,$\mu$m).}
\label{fig:C_abs-fayalite-SPH-DHS}
\end{center}
\end{figure}

\clearpage

\clearpage
\begin{table}
\caption{Information regarding observational data and blackbody curves.}
\small
\label{tab:lightcurve}
\begin{tabular}{lccccc}
\hline
Observation & TDT     & Observation  & Visual   & $T_{\rm eff}$      & $T_{dust}$\\
Number     & Number    & Date        & magnitude & (K) 	             & (K)\\
\hline
T~Cep1    & 26300141   & 08/05/1996  & 9.7   	 & 2703        & 575\\
T~Cep2    & 34601646   & 10/27/1996  & 9.8       & 2683        & 550\\
T~Cep3    & 42602251   & 01/15/1997  & 8.0       & 3035		   & 600\\
T~Cep4    & 51401256   & 04/13/1997  & 6.4       & 3347        & 650\\
T~Cep5    & 57501031   & 06/16/1997  & 7.7       & 3093			& 600\\
T~Cep6    & 66101436   & 09/07/1997  & 10.0      & 2644			& 550\\
T~Cep7    & 74602101   & 12/01/1997  & 10.4      & 2566			& 500\\
\hline
\end{tabular}
\end{table}

\clearpage
\begin{table}
\caption{Sources of laboratory mineral data.}
\small
\label{tab:labdata}
\begin{tabular}{l@{\hspace{1mm}}c@{\hspace{1mm}}c@{\hspace{1mm}}c@{\hspace{1mm}}c@{\hspace{1mm}}c}
\hline
Sample    & Chemical        & Designated     & Original form     & References\\
          & Composition     & by             & of lab data\\
\hline
Amorp. enstatite & $\rm MgSiO_3$  & -  & optical constants ($n,k$) & J\"{a}ger et al. (1994)\\
Amorp. alumina & $\rm Al_2O_3$    & -  & optical constants ($n,k$) & Begemann et al. (1997)\\
Cryst. spinel & $\rm MgAl_2O_4$   & -  & optical constants ($n,k$) & Fabian et al. (2001)\\
Amorp. Mg-Fe oxides & $\rm Mg_{0.1}Fe_{0.9}O$ & - & optical constants ($n,k$) & Henning et al. (1995)\\
Cryst. Olivine & $\rm (Mg,Fe)_2SiO_4$  & FoX  & absorbance ($a$)    & Pitman et al. (2010)\\
Cryst. Pyroxene & $\rm (Mg,Fe)SiO_3$   & EnX  & absorbance ($a$)	  & Hofmeister et al. in prep.\\
Cryst. alumina    & $\rm Al_2O_3$       & Cor       & optical constants ($n,k$) & Gervais   (1991)\\
Cryst. Forsterite  & $\rm Mg_2SiO_4$     & Fo100       & optical constants ($n,k$) & Mukai \& Koike (1990)\\
Cryst. Mg-rich Olivine  & $\rm Mg_{1.9}Fe_{0.1}SiO_4$ & Fo90 & optical constants ($n,k$) & Fabian et al. (2001)\\
Cryst. Fayalite    & $\rm Fe_2SiO_4$     & Fo0       & optical constants ($n,k$) & Fabian et al. (2001)\\
\hline
\end{tabular}
\end{table}

\clearpage
\begin{table}
\caption{Correlations between flux ratios for the observed spectral features.  }
\label{tab:fluxratio}
\begin{tabular}{rlcc}
\hline
\multicolumn{2}{l}{Continuum measured at $\rightarrow$} & 8.2$\mu$m &  40$\mu$m\\
\multicolumn{2}{l}{Peak Position} & \multicolumn{2}{c}{$R^2$}\\
\hline
${\rm F_{9.7}}$    & ${\rm F_{11.3}}$  & \bf 0.913 & \bf 0.958\\
${\rm F_{9.7}}$    & ${\rm F_{13.1}}$  & 0.489 & \bf 0.899\\
${\rm F_{9.7}}$    & ${\rm F_{20.0}}$  & \bf 0.576 & \bf 0.863\\
${\rm F_{9.7}}$    & ${\rm F_{32.0}}$  & 0.439 & \bf 0.579\\
\hline
${\rm F_{11.3}}$   & ${\rm F_{13.1}}$  & 0.334 & \bf 0.797\\
${\rm F_{11.3}}$   & ${\rm F_{20.0}}$  & \bf 0.633 & \bf 0.826\\
${\rm F_{11.3}}$   & ${\rm F_{32.0}}$  & \bf 0.504 & \bf 0.537\\
\hline
${\rm F_{13.1}}$   & ${\rm F_{20.0}}$  & \bf 0.545 & \bf 0.869\\
${\rm F_{13.1}}$   & ${\rm F_{32.0}}$   & \bf 0.567 & \bf 0.530\\
\hline
${\rm F_{20.0}}$   & ${\rm F_{32.0}}$  & 0.215 & 0.323\\
\hline
\end{tabular}\\

\begin{tabular}{p{6.5in}}
Coefficients of determination ($R^2$) $>$ 0.5 constitutes a correlation;\\
{\bf bold} designates to which there are strong/significant correlation.\\
\end{tabular}
\end{table}

\end{document}